\documentclass[a4paper,fleqn]{cas-sc}
\usepackage[authoryear]{natbib} 
\usepackage{caption}
\usepackage{threeparttable}
\usepackage{multirow}
\usepackage{amsmath}
\usepackage[normalem]{ulem}
\usepackage{algorithm} 
\usepackage{algorithmic}
\usepackage{booktabs}
\usepackage{clrscode3e}
\usepackage{hyperref}
\usepackage{multirow}
\usepackage{graphicx}
\usepackage{subfigure}
\usepackage{textcomp}
\usepackage{xcolor}
\usepackage{acronym}
\usepackage{amsmath}

\usepackage{amssymb}
\usepackage{amsfonts}
\usepackage{amsthm}
\usepackage{mathrsfs}
\usepackage{indentfirst}
\usepackage{cases}
\usepackage{bm}
\usepackage{color}
\usepackage{stfloats}
\usepackage{url}
\usepackage{verbatim}
\usepackage{array}
\usepackage{float}

\usepackage{acronym}

\acrodef{CDR}{Cross-domain Recommendation}
\acrodef{CSR}{Cross-domain Sequential Recommendation}
\acrodef{DA-GCN}{Domain-Aware Graph Convolutional Network}
\acrodef{GR}{Group Recommendation}
\acrodef{MDP}{Markov Decision Process}
\acrodef{RL-DF}{Reinforcement learning-enhanced Domain Filter}
\acrodef{GRU}{Gated Recurrent Unit}
\acrodef{MLP}{Multi-Layer Perceptron}
\acrodef{SAM}{Shared-Account Modeling}
\acrodef{BCR}{Basic Cross-domain Recommender}
\acrodef{UIN}{User Identification Network}
\acrodef{GNN}{Graph Neural Network}
\acrodef{GNNs}{Graph Neural Networks}
\acrodef{GCN}{Graph Convolutional Network}
\acrodef{GCNs}{Graph Convolutional Networks}
\acrodef{SCRM}{Self-attention-based Cross-domain Recommendation Machine}
\acrodef{SR}{Sequential Recommendation}
\acrodef{RNNs}{Recurrent Neural Networks}
\acrodef{RNN}{Recurrent Neural Network}
\acrodef{CNN}{Convolutional Neural Network}
\acrodef{CNNs}{Convolutional Neural Networks}
\acrodef{MRR}{Mean Reciprocal Rank}
\acrodef{CDS}{Cross-Domain Sequence}
\acrodef{CSR}{Cross-domain Sequential Recommendation}
\acrodef{CL}{Contrastive Learning}
\acrodef{GCL}{Graph Contrastive Learning}
\acrodef{SGL}{Self-supervised Graph Learning}
\acrodef{SOTA}{state-of-the-art}

\acrodef{PPCDR}{Privacy-Preserving Cross-Domain Recommendation}
\acrodef{FGSAT}{Fine-Grained
Semantic Adaptation and Transfer}
\acrodef{KG}{knowledge graph}

\acrodef{DAL}{Domain-Adversarial Learning}
\acrodef{RSA}{Representation Space Alignment}
\acrodef{FL}{Federated learning}

\begin{document}



\shorttitle{FedCRF: A Federated Cross-domain Recommendation Method}


\title [mode = title]{FedCRF: A Federated Cross-domain Recommendation Method with Semantic-driven Deep Knowledge Fusion}                      
        



%

\author[1]{Lei Guo}
\ead{leiguo.cs@gmail.com}

\author[1]{Ting Yang}
\ead{dcongtou@gmail.com}

\author[2]{Xu Yu}
\ead{yuxu0532@upc.edu.cn}
\credit{Resources, Validation, Data curation}

\author[3,4]{Xiaohui Han}
\ead{xiaohhan@gmail.com}
\cormark[1]

\author[5]{Guiyuan Jiang}
\ead{jiangguiyuan@ouc.edu.cn}

\author[6]{Hui Liu}
\ead{liuh_lh@sdufe.edu.cn}
\cormark[2]

\cortext[cor1]{Corresponding author}
\cortext[cor2]{Corresponding author}

\affiliation[1]{organization={School of Computer and Artificial Intelligence, Shandong Normal University},
    city={Jinan},
    postcode={250358}, 
    country={China}}

\affiliation[2]{organization={Qingdao Institute of Software, China University of Petroleum (East China)},
    city={Qingdao},
    postcode={266580}, 
    country={China}}

\affiliation[3]{organization={Key Laboratory of Computing Power Network and Information Security, Ministry of Education, Shandong Computer Science Center (National Supercomputer Center in Jinan), Qilu University of Technology (Shandong Academy of Sciences)},
    city={Jinan},
    postcode={250000}, 
    country={China}}

\affiliation[4]{organization={Shandong Provincial Key Laboratory of Industrial Network and Information System Security, Shandong Fundamental Research Center for Computer Science},
    city={Jinan},
    postcode={250000}, 
    country={China}}

\affiliation[5]{organization={School of Computer Science and Technology, Ocean University of China},
    city={Qingdao},
    postcode={266100}, 
    country={China}}

\affiliation[6]{organization={School of Computing and Artificial Intelligence, Shandong University of Finance and Economics},
    city={Jinan},
    postcode={250014}, 
    country={China}}

\begin{abstract}
As user behavior data becomes increasingly scattered across different platforms, achieving cross-domain knowledge fusion while preserving privacy has become a critical issue in recommender systems. Existing \ac{PPCDR} methods usually rely on overlapping users or items as the bridge, making them inapplicable to non-overlapping scenarios.
They also suffer from limitations in the collaborative modeling of global and local semantics.
To this end, this paper proposes a \textit{Federated Cross-domain Recommendation method with deep knowledge Fusion} (FedCRF).
Using textual semantics as a cross-domain bridge, FedCRF achieves cross-domain knowledge transfer via federated semantic learning under the non-overlapping scenario. Specifically, FedCRF constructs global semantic clusters on the server side to extract shared semantic information, and designs a \ac{FGSAT} module on the client side to dynamically adapt to local data distributions and alleviate cross-domain distribution shift. Meanwhile, it builds a semantic graph based on textual features to learn representations that integrate both structural and semantic information, and introduces contrastive learning constraints between global and local semantic representations to enhance semantic consistency and promote deep knowledge fusion.
In this framework, only item semantic representations are shared, while user interaction data remains locally stored, effectively mitigating privacy leakage risks.
Experimental results on multiple real-world datasets show that FedCRF significantly outperforms existing methods in terms of Recall@20 and NDCG@20, validating its effectiveness and superiority in non-overlapping cross-domain recommendation scenarios.

\end{abstract}


\maketitle

\section{Introduction \label{sec:introduction}}
In recent years, the growing diversity of user behavior across different systems has driven researchers in recommender systems to explore methods for efficiently integrating knowledge from heterogeneous sources to enhance recommendation performance. This has led to the rise of \ac{CDR}, which seeks to address issues like data sparsity and the cold-start problem in the target domain by leveraging valuable knowledge from other domains. As a result, \ac{CDR} has garnered significant attention in both academic research and industry applications~\citep{li2022scdgn,liu2022CFAA,zhu2022PTUPCDR,guo2022TiDA-GCN}.

Traditional \ac{CDR} methods typically rely on overlapping users between domains to enable the transfer of user preferences~\citep{zhu2022PTUPCDR,zhu2020GADTCDR}. In practical applications, the demand for cross-domain collaboration is growing between different independent entities, such as e-commerce platforms and fresh delivery apps, or offline supermarkets and online shopping guides. These organizations want to leverage each other's knowledge to optimize recommendations and expand user coverage. However, due to privacy regulations like GDPR~\citep{protection2018GDPR} and commercial confidentiality restrictions, they cannot share sensitive information such as user IDs and interaction records. More importantly, the users and products of these organizations often do not overlap, making traditional \ac{CDR} methods that rely on overlapping entities ineffective. The non-overlapping privacy-preserving cross-domain recommendation scenario we focus on aims to solve this issue: enabling cross-domain knowledge transfer without sharing sensitive data or relying on overlapping entities. This approach ensures compliance while enriching isolated domains with knowledge, bridging the gap between "collaboration needs" and "privacy constraints".

To overcome this challenge, researchers have developed non-overlapping cross-domain recommendation methods~\citep{li2022gromov,li2023SRTrans,cao2022DisenCDR,li2022scdgn,liu2022CFAA,wang2019recsys} to facilitate effective knowledge transfer in settings where users are disjoint. These methods enable cross-domain knowledge transfer by learning latent structural or semantic relationships across domains (e.g., representation disentanglement, structural modeling, or distribution alignment), thereby achieving partial fusion at the representation level without directly sharing user data. While these methods show promising results in non-overlapping scenarios, they often depend on centralized training or the sharing of interaction data across domains, which may violate data compliance and user privacy regulations.
This is especially true in situations where user behavior data cannot be transferred outside its local domain, severely limiting the practical applicability of existing approaches.

\ac{PPCDR} has emerged as a crucial research topic. Current approaches to this task typically leverage federated learning frameworks, enabling multiple participants to collaboratively train models without sharing raw interaction data, thereby improving recommendation performance~\citep{liao2023ppgencdr,liu2023ppmdr,meihan2022fedcdr,lin2024P2DTR, wang2024P2M2-CDR, wang2025fedpcl,jia2025perfedgt,yan2022fedcdr}. 
For instance,~\citet{meihan2022fedcdr} developed FedCDR with dual-module structure: a personal module for local feature processing and a transfer module for cross-domain knowledge sharing, preventing negative transfer.~\citet{chen2022pricdr} proposed PriCDR, which introduces a differential privacy mechanism in the source-domain rating matrix publishing stage. It achieves privacy protection by perturbing the data, and then performs cross-domain recommendation modeling based on the published data. 

However, existing \ac{PPCDR} methods still have the following limitations: 1) They often depend on overlapping users or items as "bridges" for cross-domain knowledge transfer, which complicates the task of fully utilizing cross-domain information while safeguarding data privacy. In real-world datasets, the overlap of users is typically very limited, making it challenging to achieve both effective knowledge transfer and privacy protection simultaneously. As a result, achieving truly non-overlapping cross-domain knowledge transfer while ensuring privacy remains largely unresolved. 2) Previous methods cannot integrate the local and global domain information well, due to the distribution shift between them. Typical non-overlapping cross-domain methods like FFMSR~\citep{lu2025FFMSR} use server-side static clustering to aggregate item information across domains and build a unified global semantic representation. However, the global representation generated via server-side clustering can be viewed as a highly compressed abstraction of cross-domain semantics, mainly capturing domain-agnostic high-level features. In contrast, the local clients tend to model highly granular feature representations (e.g., needing to distinguish between "hard sci-fi" and "romantic sci-fi" rather than relying on coarse "sci-fi" labels). Therefore, there remains a clear semantic distribution gap between the global and local representations, leading to difficulties in cross-domain alignment and degrading model performance. Consequently, there is a need for intelligent systems that can dynamically balance global consistency with local relevance. 3) They primarily concentrate on the ID modality to represent items, while the importance of the text modality is often overlooked~\citep{yan2022fedcdr,zhang2024fedhcdr}. Many approaches~\citep{wang2025fedpcl} merely integrate the two modalities superficially, using techniques such as addition or concatenation, without achieving deep integration.

To address the above limitations, we propose a \textit{Federated Cross-domain Recommendation method with
deep knowledge Fusion} (FedCRF) that unifies global semantic consensus, local fine-grained semantics, and ID-based interaction features. 
Specifically, to address \textbf{Limitation 1}, we leverage the universality of natural language to bridge disjoint domains.
In the semantic space, semantically similar textual content is mapped to similar vector representations.
Building on this intuition, we map items from diverse domains into a shared semantic space and employ server-side clustering to create global semantic centers that capture cross-domain consensus. This approach fosters semantic-level associations rather than entity-level dependencies, enabling a seamless knowledge transfer across non-overlapping domains based purely on semantic understanding. 
To address \textbf{Limitation 2}, we introduce the \ac{FGSAT} module. This module leverages the global cluster centers from the server as semantic seeds and dynamically updates the semantic cluster centers with local fine-grained textual features. Through an attention mechanism, it dynamically integrates the global semantic structures with the local, detailed features, facilitating knowledge remapping within the local semantic context.
In Stage 2, we further refine the pre-learned domain knowledge from the previous stage through bidirectional contrastive learning. This approach creates dual-view representations and uses contrastive loss to automatically balance the contribution of global and local features. To deal with \textbf{Limitation 3}, we perform semantic fusion at the modality level, as relying solely on a single modality or simple concatenation fails to capture item-specific variations. To address this, we construct semantic graphs from globally pre-trained and locally learned representations, where each graph injects inter-item relationships into item IDs through graph convolution operations, generating fused text-ID representations. Ultimately, semantic and ID representations are merged into a unified multi-modal representation, achieving significant performance improvements.


\textbf{The motivation and main contributions are summarized as follows:}

In this work, we study the \ac{PPCDR} task under the non-overlapping scenario by proposing a semantic-driven knowledge fusion method to address the global information fusion and the limited integration of ID and semantic features.

To conduct domain knowledge transfer under the non-overlapping scenario, we resort to the federated semantic learning. Specifically, we do not align domains by specific users or items. Instead, we fuse them at the semantic level, where only item information is uploaded to the server, and the user information is only used within the local client.
The GNN-based method is exploited as the backbone of our method to learn user preferences and item representations.
From the experimental results in Table~\ref{tab:Privacy_Risk}, we can find that user information can be protected to a large extent.

To transfer knowledge across domains, we then upload the item representations from local clients to the server for knowledge fusion, where a static clustering method is utilized for this purpose.
However, we find that static clustering can only provide coarse-grained semantic fusion (as shown in Figures~\ref{fig:cross_domain_tsne}(a),~\ref{fig:cross_domain_tsne}(b) and~\ref{fig:cross_domain_tsne}(c)). It cannot well capture the distribution of the local data.
To deal with this, we further conduct clustering on item representations with an improved clustering method, where the global domain information is introduced as a guidance (the visualization results of this process can be seen in Figures~\ref{fig:cross_domain_tsne}(d),~\ref{fig:cross_domain_tsne}(e) and~\ref{fig:cross_domain_tsne}(f)).

To further leverage the global domain knowledge from the server to update local clients, we then design a contrastive learning method based on semantic graphs constructed by local semantic information and pre-learned semantic information during Stage 2 (fine-tuning on local clients with semantic fusion) (as shown in Figure~\ref{fig:stage2}).
By doing this, we can further transfer the pre-learned domain knowledge to specific domains to enhance its recommendation results.

In our method, the attention mechanism is used to adaptively weight and fuse semantic information at different levels (as shown in Figure~\ref{fig:FGSAT}). Specifically, by assigning dynamic weights to the enhanced item semantic representations, global semantic centers, and fine-grained semantic centers, the model can selectively emphasize more discriminative semantic sources according to the semantic characteristics of the current item. This mechanism effectively alleviates the bias caused by single semantic information and improves the flexibility and expressiveness of multi-source semantic fusion.

We conduct extensive experiments on three pairs of domains to evaluate our method.
From the experimental results in Table~\ref{tab:results} and Table~\ref{tab:cross_platform}, we can find that our method can perform best on all the evaluation metrics. These results demonstrate the effectiveness of our method on the privacy-preserving cross-domain recommendation task.

\section{Related Work \label{sec:relatedwork}}
This section reviews related work on \ac{CDR}, Federated Cross-Domain Recommendation, and Semantic Representation Learning for \ac{CDR}.

\subsection{PPCDR}

\ac{CDR} aims to use knowledge from a source domain to alleviate data sparsity and cold-start problems in a target domain. Its methods have evolved from relying heavily on overlapping entities to modeling cross-domain knowledge in non-overlapping scenarios. Early methods usually rely on explicit overlapping entities for knowledge transfer. For example, EMCDR~\citep{man2017EMCDR} learns an embedding mapping function across domains to achieve representation alignment, while GA-DTCDR~\citep{zhu2020GADTCDR} introduces graph-based modeling and alignment mechanisms to enhance cross-domain relational representation at the structural level. Later methods relax the requirement of overlap and improve transfer performance through semantic enhancement and representation alignment. SEAGULL~\citep{zhao2023SEAGULL} uses semantic enhancement and contrastive learning to reduce domain shift, while CVPM~\citep{zhao2025CVPM} applies projection learning and collaborative modeling to enable transfer in weakly shared spaces. In more weak or even non-overlapping settings, KR-CDR~\citep{huang2024KR-CDR} builds a unified semantic space using knowledge graphs to support structured transfer, CD2CDR~\citep{zhu2025CD2CDR} mines latent cross-domain relations through bidirectional mapping or dual learning, and ARISEN~\citep{zhao2023ARISEN} combines time-aligned behavior sequences with semantic decomposition to achieve fine-grained cross-domain knowledge transfer without explicit overlaps. However, most of these methods are still based on centralized learning paradigms, assuming that data from different domains can be collected and jointly trained. As a result, they generally lack explicit privacy-preserving mechanisms, such as federated learning or differential privacy techniques.

Recently, \ac{PPCDR} has emerged as a promising research direction, aiming to achieve secure cross-domain knowledge transfer without sharing raw user data. According to different privacy-preserving mechanisms, existing methods can be roughly divided into two categories. The first category is distributed collaborative modeling based on federated learning~\citep{meihan2022fedcdr,zhang2024fedhcdr,zheng2024fedgkd,wang2025fedpcl,zhong2024scfl,bao2026pFedSG,gamiz2025prot,jia2025perfedgt}. These approaches enable joint training among multiple parties without moving local data out of each domain, achieving collaborative modeling of cross-domain information and global knowledge fusion. For example,~\citet{bao2026pFedSG} introduced a semi-decentralized architecture to enable device-to-device collaboration and improved recommendation performance with a fine-grained personalized aggregation mechanism.~\citet{jia2025perfedgt} addressed the scale heterogeneity problem in federated graph data by introducing a graph transformer and a hypernetwork, enabling personalized modeling and stable training across different clients. In addition,~\citet{zhong2024scfl} improved personalized POI recommendation in federated settings by modeling spatio-temporal consistency of trajectories and introducing an edge-based collaborative aggregation mechanism. These methods achieve cross-domain or cross-client knowledge sharing while ensuring that raw data remains local. The second category is cross-domain recommendation based on knowledge transfer. These methods perform cross-domain knowledge transfer under privacy constraints by aligning, mapping, or generating user/item representations. For example,~\citet{liao2023ppgencdr} modeled the interaction distribution in the source domain using a generative adversarial network and released synthetic data under differential privacy mechanisms. This enables cross-domain knowledge transfer without sharing raw interaction data, while improving recommendation performance.

Although the above methods have achieved significant progress in \ac{PPCDR}, most existing approaches still rely on overlapping users or shared entities as bridges for knowledge transfer, which limits their applicability in real-world scenarios. In fully non-overlapping settings with strict privacy isolation, how to achieve efficient cross-domain knowledge transfer and high recommendation performance under strong privacy constraints remains a key challenge in federated cross-domain recommendation.

\subsection{Semantic Modeling for CDR}

In \ac{CDR}, semantic information is widely used to support cross-domain knowledge transfer and alleviate data sparsity caused by non-overlapping users or items. To this end, textual and visual modalities are often used as auxiliary inputs in representation learning to improve cross-domain modeling ability. Existing studies show a clear evolution from simple feature augmentation to semantic structure modeling. Early methods mainly use text as auxiliary features for representation learning. For example, CD2CDR~\citep{zhu2025CD2CDR} improves user preference modeling by incorporating user reviews and item textual content, but it does not explicitly model cross-domain semantic relationships. Later, methods such as ARISEN~\citep{zhao2023ARISEN} and SEAGULL~\citep{zhao2023SEAGULL} further exploit semantic structures from text and images to reduce cross-domain distribution gaps. In addition, KR-CDR~\citep{huang2024KR-CDR} introduces knowledge graphs to enhance structured representations, while MITrans~\citep{li2024MITrans} improves cross-domain alignment through mutual information modeling, which strengthens performance in non-overlapping scenarios.

Under privacy constraints, recent studies have begun to introduce semantic representation learning into federated learning and cross-domain recommendation to enable cross-domain knowledge sharing without accessing raw data. However, most existing methods still rely mainly on ID-based representations, and the use of textual or other semantic modalities remains relatively shallow, with limited integration between semantic information and federated optimization. In this context, some methods attempt to use semantic information for privacy-preserving cross-domain transfer. For example, PFCR~\citep{guo2024PFCR} applies vector quantization techniques to discretize item description texts and combines the resulting representations with joint representation learning for cross-domain knowledge transfer. However, it compresses continuous semantic spaces into discrete codes, which leads to loss of fine-grained semantic information. It also depends on codebook updates, which may introduce representation drift and reduce semantic expressiveness.
Further, FFMSR~\citep{lu2025FFMSR} works in privacy-preserving settings without overlapping users or items. It builds global semantic prototypes on the server through semantic clustering for cross-domain alignment and transfer. However, these prototypes are based on static clustering results and cannot adapt to local client distributions. As a result, semantic alignment is mainly a one-time global matching process and cannot capture dynamic local semantic patterns.

Different from these methods, our approach introduces a semantic center re-estimation mechanism within a global semantic clustering framework. This allows semantic prototypes to be dynamically updated according to local data distributions, enabling joint modeling of global semantic structure and local characteristics. In this way, semantic alignment is extended from static prototype matching to dynamic semantic reconstruction, improving fine-grained cross-domain knowledge transfer.

\subsection{Difference}

In cross-domain recommendation research, existing methods generally follow core steps such as semantic modeling, representation learning, and cross-domain alignment, resulting in certain similarities in their overall frameworks. However, significant differences still exist across methods in terms of task settings, modeling targets, semantic representation forms, cross-domain knowledge transfer mechanisms, and privacy constraints. From these key dimensions, we conduct a systematic comparative analysis between the proposed method and representative works. Overall, under the non-overlapping user setting, our method takes continuous semantic representations as a bridge and achieves effective transfer and fusion of cross-domain knowledge through federated semantic learning and dynamic semantic modeling.

The differences between our method and existing works are summarized as follows:

(1) Our method has significant differences with the FFMSR method~\citep{lu2025FFMSR}
in the following aspects:
1) In our method, we focus on devising models to conduct the item prediction task from user-item pairs. Our task is to model users' preferences on specific items.
Different from us, FFMSR aims at modeling users' sequential preferences by predicting the next item that the user will interact with.
2) In user preference modeling, FFMSR exploits sequential-based methods, such as SASRec~\citep{kang2018SASRec}. 
But this kind of method can only model users' sequential preferences, and the high-order connections among users and items are largely ignored.
Hence, we resort to the graph-based method to improve it for information fusion and domain adaptation.
3) Although FFMSR also uses semantic information as a cross-domain bridge, it maps textual semantics to discrete cluster centers through clustering, which compresses the semantic representation space and easily leads to the loss of fine-grained information. In contrast, our method models based on continuous textual semantic representations to achieve finer-grained cross-domain association.
4) To connect disjoint domains in federated learning, FFMSR exploits a static clustering method at the server to fuse the global item information from different domains.
But we find that the resulting global representation has a large distribution gap with the local domain representation, resulting in the distribution shift issue.
To resolve this, we further devise an improved clustering method in local clients, which is guided by the global information. From the experimental results in Figure~\ref{fig:cross_domain_tsne}, we find that our method can better integrate global and local domain information.

(2) Our method is significantly different from the UniCDR method~\citep{cao2023UniCDR} 
in the following aspects: 1) Our research target is to conduct privacy-preserving cross-domain recommendations, while UniCDR tends to devise a unified cross-domain model for different domain settings.
But their method relies on joint modeling of multi-domain data, which requires integrating and sharing user interaction information across domains, and cannot protect domain privacy.
2) To conduct domain knowledge transfer, UniCDR needs overlapping users or items as a bridge, while our method studies cross-domain recommendation under the totally non-overlapping scenario.
In contrast to UniCDR, which focuses on learning domain-shared information via overlapping users or items, we align domains at the semantic level, not at specific users or items.
To achieve this, we devise a global-aware semantic clustering method to fuse the domain-shared information into local domains for knowledge transfer.

(3) Our method is significantly different from the FedPCL-CDR method~\citep{wang2025fedpcl}  
in the following aspects: 1) 
FedPCL-CDR relies on overlapping users for cross-domain transfer and uploads differential prototypes together with related user sets, which brings privacy risks. Meanwhile, its local differential privacy mechanism faces a trade-off between recommendation performance and privacy protection. In contrast, our method only uploads item embeddings under the non-overlapping setting, and user data is always kept locally, which inherently reduces privacy risks. 
2) 
FedPCL-CDR constructs cross-domain prototypes based on overlapping users and achieves knowledge transfer through prototype-level alignment, which is a static prototype-driven mechanism limited by prototype expressiveness and data distribution.
In contrast, starting from the semantic representation space, our method dynamically re-estimates semantic centers under global clustering constraints to form secondary semantic centers, and models item relations with semantic graphs, thus achieving deep fusion of global and local information.
3) 
FedPCL-CDR constructs contrastive objectives based on user representations and prototypes, whose contrastive granularity is restricted by prototype representations. In contrast, our method builds a semantic-aware item graph based on textual features to learn representations fused with structural semantic information, and imposes contrastive constraints between representations from global and local semantic views, thereby achieving finer-grained cross-domain alignment and more sufficient knowledge fusion.

(4) Our method is significantly different from the FedDCSR method~\citep{zhang2024feddcsr}  
in the following aspects: 1) In our method, we focus on devising models to conduct the item prediction task from user-item pairs. Our task is to model users' preferences on specific items.
Different from us, FedDCSR aims at modeling users' sequential preferences by predicting the next item that she will interact. 2) FedDCSR achieves cross-domain transfer by weighted aggregation of domain-shared representations and model parameters, which is essentially explicit fusion at the user representation level. In contrast, our method starts from item semantics, performs cross-domain clustering on item representations, and dynamically re-estimates semantic centers under global clustering constraints. Meanwhile, it models item relations via semantic graphs to achieve deep fusion of global and local information. This mechanism does not require sharing user-level information, keeping user interaction data local and effectively reducing privacy leakage risks.
3) FedDCSR constructs positive and negative samples by data augmentation on user sequences, which belongs to sample perturbation-based contrastive learning. In contrast, our method imposes contrastive constraints between global and local fused representations. By aligning their semantic consistency, it achieves finer-grained cross-domain alignment and deeper knowledge fusion.

\section{Method \label{sec:method}}
This chapter systematically elaborates on the proposed FedCRF. The chapter is organized as follows:  Section 3.1 defines the problem formulation and explains key notations. Section 3.2 presents the overall architecture and workflow of FedCRF. Section 3.3 describes the learning process of semantic and ID representations that serve as fundamental inputs to the model. Sections 3.4 and 3.5 introduce the server-side clustering mechanism and the client-side fine-grained semantic adaptation and transfer module during the federated pre-training phase. Finally, Section 3.6 comprehensively analyzes the local fine-tuning stage, covering graph-based learning, contrastive fusion mechanisms, and the final recommendation prediction.
\begin{figure*}
    \centering
    \includegraphics[width=15cm]{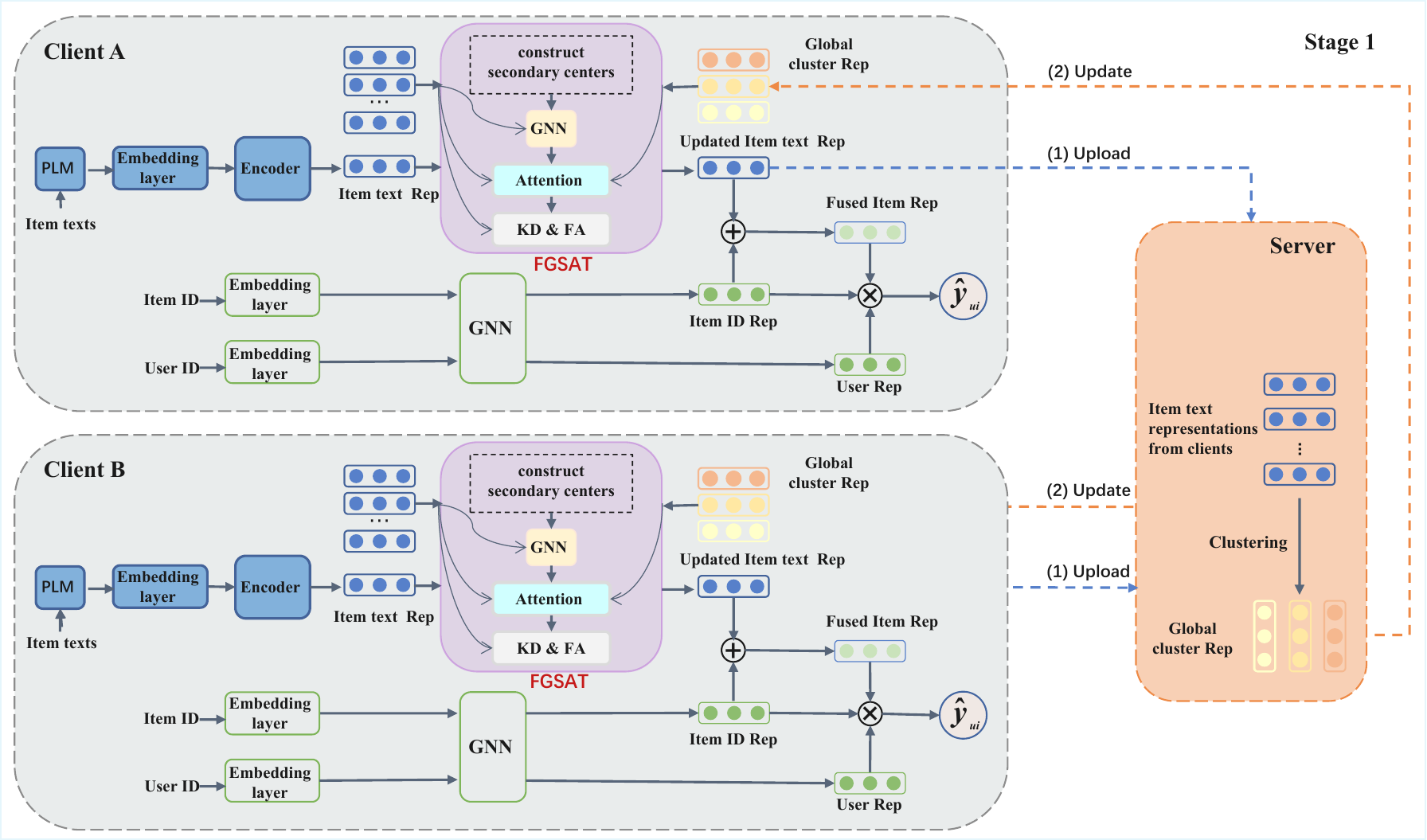}
    \caption{The workflow of FedCRF in Stage 1: 1) Each client uploads its locally encoded item text representations to the server. 2) The server performs global semantic clustering based on the aggregated representations. 3) The clustering results are distributed to each client. 4) Under fixed global semantic assignment constraints, the client regularizes and re-estimates semantic centers through the \ac{FGSAT} module, and trains by fusing multi-source semantic information. 5) It fuses text representations and ID representations to complete prediction and model optimization. 6) After training, the updated semantic representations are saved for subsequent tasks.}
    \label{fig:stage1}
\end{figure*}

\begin{figure}
    \centering\includegraphics[width=0.8\linewidth]{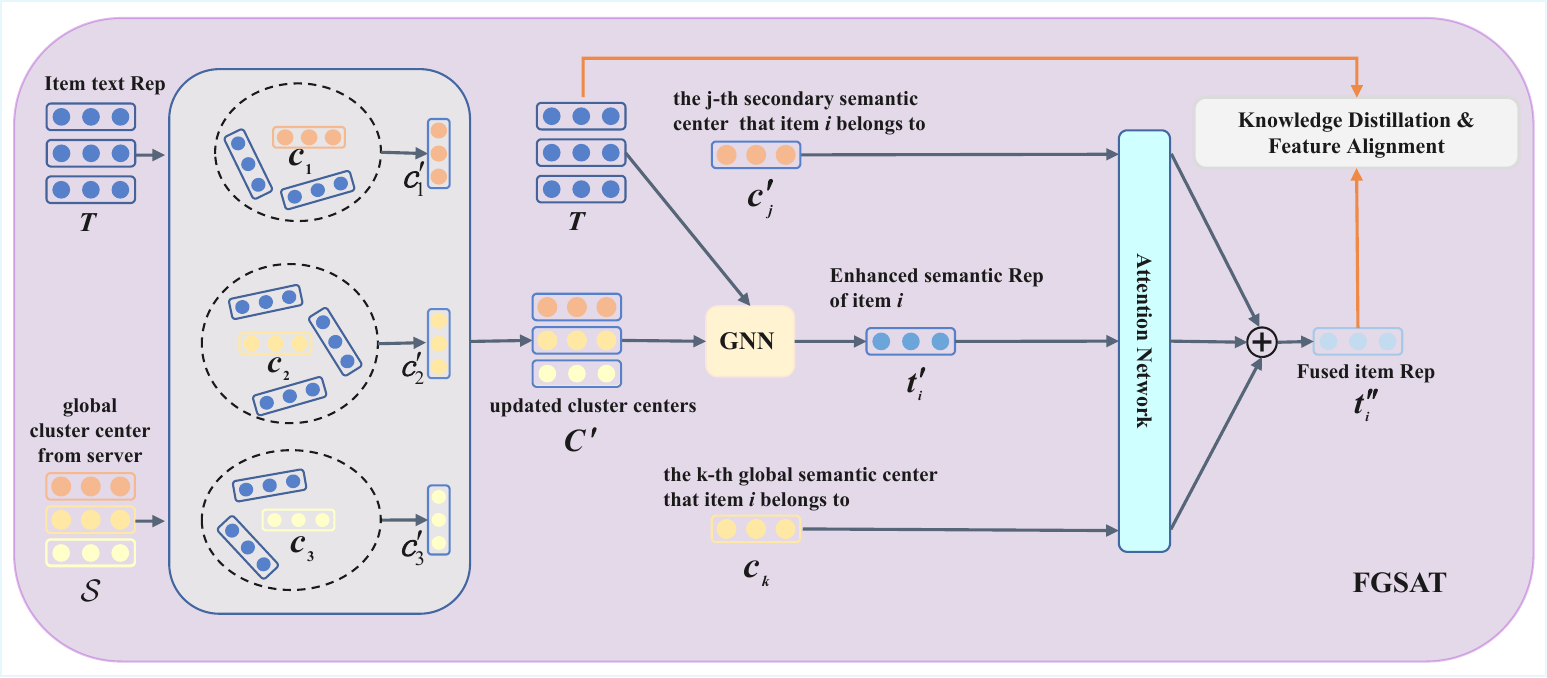}
    \caption{The workflow of the \ac{FGSAT} module in Stage 1 of FedCRF framework. First, secondary semantic centers $\boldsymbol{C}'$ are constructed based on global clusters $\mathcal{S}$ and local text representations $\boldsymbol{T}$, and a graph structure is built using $\boldsymbol{T}$ and $\boldsymbol{C}'$. Next, \ac{GNN} propagates semantics over the graph. Then, a dynamic fusion strategy based on an attention mechanism integrates multi-source semantic information from the global clusters $\boldsymbol{c}_k$, local secondary centers $\boldsymbol{c}_j'$, and the transferred item representations $\boldsymbol{t}'_i$. Finally, knowledge distillation and feature alignment are applied to ensure semantic consistency.}
    \label{fig:FGSAT}
\end{figure}

\subsection{Preliminaries}
Suppose that we have two domains $\mathrm{A}$ and $\mathrm{B}$. Let $\mathcal{U}_A$ and $\mathcal{U}_B$ denote the user sets, $\mathcal{I}_A$ and $\mathcal{I}_B$ denote the item sets in domains $\mathrm{A}$ and $\mathrm{B}$, respectively. Each item $i^A \in \mathcal{I}_A$ (or $i^B \in \mathcal{I}_B$) is identified by a unique item ID and associated with a descriptive text (e.g., its title, description, and brand). Taking domain $\mathrm{A}$ as an example, each user $u \in \mathcal{U}_A$ is associated with a set of items $\mathcal{I}_u$ with positive feedback, indicating a preference score of $y_{ui} = 1$ for $i \in \mathcal{I}_u$. Besides user-item interactions, textual features are used as content information of items. We utilize Sentence Transformers~\citep{reimers2019sentence} to encode the concatenated descriptive text of item $i$ into a 1024-dimensional sentence embedding. This is denoted as the textual feature $\boldsymbol{x}^t_i$.

Under the non-overlapping cross-domain recommendation setting, there is no overlap in users or items across domains. Our goal is to leverage both textual information and unique IDs to accurately predict users' preferences by ranking items for each user according to predicted preference scores $\hat{y}_{ui}$.

\subsection{Overview of FedCRF}
\begin{figure}
    \centering
    \includegraphics[width=0.8\linewidth]{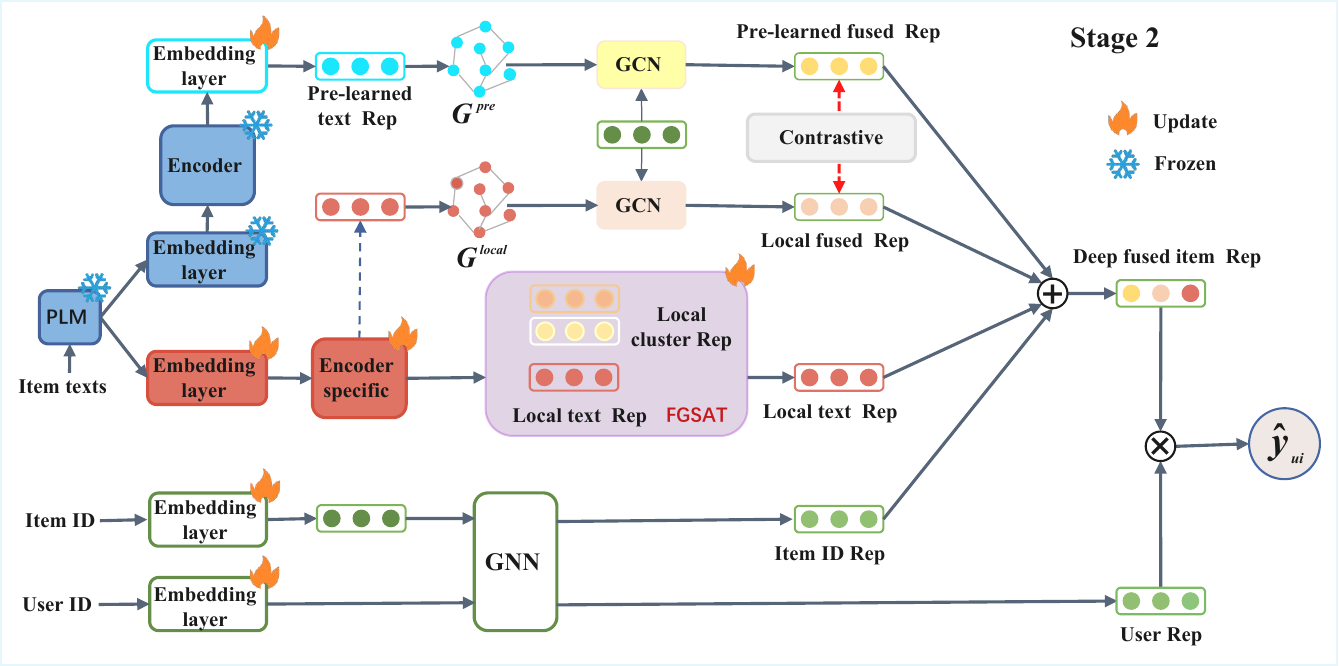}
    \caption{The workflow of FedCRF in Stage 2. 1) The client loads the pre-trained semantic representations obtained in Stage 1 and the representations generated by the local specific encoder, and constructs the pre-trained semantic graph and the local semantic graph respectively. 2) It performs representation learning via graph convolution on the two semantic graphs together with ID embeddings, obtaining the pre-trained fused representation and the local fused representation. 3) A contrastive learning constraint is introduced between the pre-trained and local fused representations. 4) It fuses multi-source representations and completes prediction and recommendation combined with user representations.}
    \label{fig:stage2}
\end{figure}

As illustrated in Figure~\ref{fig:stage1} and Figure~\ref{fig:stage2}, the proposed framework adopts a two-stage federated learning pipeline to address semantic modeling challenges under non-overlapping and privacy-preserving settings.

\textbf{Stage 1: Federated Pre-training.}  
1) First, each client encodes local item texts and uploads the generated textual representations to the server. 2) Subsequently, the server performs global semantic clustering based on the representations from all clients to derive global semantic centers capturing cross-domain shared semantics, after which the results are distributed back to each client. 3) Upon receiving the global semantic centers, each client activates the \ac{FGSAT} module (as shown in Figure~\ref{fig:FGSAT}). This module regularizes and re-estimates the semantic centers under fixed global semantic assignment constraints, allowing them to adapt to the local data distribution. This process alleviates the discrepancy between global semantics and local distributions, leading to finer-grained semantic modeling and improved cross-domain alignment. Furthermore, the \ac{FGSAT} module fuses multi-source semantic information via an attention mechanism, and introduces knowledge distillation and feature alignment constraints to facilitate effective transfer and consistency learning of cross-domain shared semantics. 4) On this basis, each client fuses textual semantic representations and ID embeddings, predicts user–item interactions, and computes the recommendation loss to update model parameters. 5) After completing this stage, each client uses the updated encoder to generate enhanced semantic representations and stores them for subsequent semantic learning and recommendation tasks.

\textbf{Stage 2: Fine-tuning on Local Clients with Semantic Fusion.}  
The objective of this stage is to achieve deep semantic fusion and local adaptive optimization based on federated pre-trained knowledge. 1) First, each client loads the pre-trained semantic representations obtained in Stage 1 as well as the representations generated by the local specific encoder, and constructs a pre-trained semantic graph and a local semantic graph, respectively. 2) Subsequently, graph convolution operations are performed on both semantic graphs to model inter-item structural relationships and inject such structural information into item ID representations, thereby generating fused representations that integrate textual semantics and ID information from different perspectives, avoiding the insufficient expressiveness caused by single-modality modeling or simple feature concatenation. 3) Based on the obtained pre-learned fused representation and local fused representation, bidirectional contrastive learning constraints are introduced to achieve finer-grained cross-domain alignment and deeper semantic fusion. 4) Finally, multi-source fused representations are integrated with user representations for the final user–item interaction prediction and recommendation task optimization.

This two-stage design of “pre-train then fine-tune locally” not only effectively mitigates potential negative transfer risks in the federated stage, but also promotes deeper semantic fusion and local adaptation, thereby improving the model's robustness, learning efficiency, and interpretability.

\subsection{Semantic and ID Representations}
\textbf{Semantic Representation.}
Our core objective is to learn high-quality semantic representations. In contrast, Large Language Models (LLMs) are mainly designed for open-domain text generation and complex reasoning, which are not fully aligned with the representation learning objective in this work. Therefore, we adopt Pre-trained Language Models (PLMs) for text encoding, specifically using Sentence-Transformers. For each item $i$, we concatenate its title, description, and brand, and feed them into the encoder to obtain a $1024$-dimensional sentence embedding $\boldsymbol{x}^t_i$:
\begin{equation}
\boldsymbol{x}^t_i = \operatorname{PLM}(\operatorname{Concat}(title_i, description_i, brand_i)).
\end{equation}

All item representations are organized into an embedding matrix \( \boldsymbol{X} \in \mathbb{R}^{|\mathcal{I}_d| \times 1024} \), where each row corresponds to the semantic embedding of an item and $\mathcal{I}_d$ denotes the item set in domain $d$. Considering that the initial semantic embeddings are high-dimensional and lack domain-specific information, we introduce a domain text encoder $\phi_{\text{emb}}$ for each domain to learn more compact and discriminative item representations. The final item embedding $\boldsymbol{t}_i$ is defined as:
\begin{equation}
\boldsymbol{t}_i = \phi_{\text{emb}}(\boldsymbol{x}^t_i),
\end{equation}
where $\phi_{\text{emb}}$ maps the initial semantic embedding into a higher-level representation space that incorporates domain-specific features. This representation preserves the textual semantics of items while enhancing domain-related characteristics, thereby improving the model's representation capability. The resulting embedding $\boldsymbol{t}_i$ will be uploaded to the server for subsequent cross-domain semantic fusion.

\textbf{User and Item ID Representation.}
For the ID embeddings of users and items, we design an ID encoder to generate corresponding embeddings based on the unique ID index of each entity. This is implemented through an ID embedding matrix $\boldsymbol{E} \in \mathbb{R}^{|\mathcal{I}_d| \times d_t}$, where $d_t$ represents the dimension of the embeddings. Each row $\boldsymbol{e}_i$ in the matrix corresponds to the ID embedding of item $i$. Similarly, the user embedding matrix $\boldsymbol{U} \in \mathbb{R}^{|\mathcal{U}_d| \times d_t}$ is used to generate embeddings for each user, where $\mathcal{U}_d$ denotes the  user set in domain $d$. Each row $\boldsymbol{e}_u$ in the matrix corresponds to the ID embedding of user $u$.

Subsequently, these user and item ID embeddings are fed into the Graph Neural Network (GNN) to further aggregate and learn the complex interaction relationships between users and items, yielding the final user and item ID representations $\boldsymbol{x}_u, \boldsymbol{x}^{id}_i \in \mathbb{R}^{d_t}$.

\subsection{Global Cluster}
To align disjoint domains within the semantic space, we employ a clustering approach. Upon receiving item text representations from various clients, we perform K-means algorithm as in~\citep{lu2025FFMSR} to capture shared semantic structures and bridge different domains. Additionally, to address the issue of inconsistent numerical feature distributions across clients in federated learning environments, we introduce a Domain-Aware Normalization module~\citep{chang2019domain}. This module standardizes feature scales, improving clustering performance.

\subsubsection{Domain-Aware Normalization}
To improve clustering performance, the domain-aware normalization module maintains a set of statistical parameters for each domain, including the mean and variance, which are used to unify feature scales and reduce distribution shifts, thereby improving the separability and stability of multi-domain clustering. Specifically, for each domain, we maintain a running mean $\boldsymbol{\mu}$ and a running variance $\boldsymbol{\sigma}^2$. Given a feature vector $\boldsymbol{t}_i$ from domain $\mathrm{A}$ or $\mathrm{B}$, the normalization process is defined as:
\begin{equation}
\widetilde{\boldsymbol{t}}_i = \frac{\boldsymbol{t}_i - \boldsymbol{\mu}}{\sqrt{\boldsymbol{\sigma}^2 + \epsilon}},
\end{equation}
where $\epsilon$ is a small constant added to avoid division by zero. 
This operation ensures that feature values are standardized across dimensions, thereby reducing scale differences between domains.

During training, the running statistics are updated by the statistics of the current mini-batch:
\begin{align}
\boldsymbol{\mu} &= \rho \boldsymbol{\mu} + (1 - \rho)\,\bar{\boldsymbol{t}}, \\
\boldsymbol{\sigma}^2 &= \rho \boldsymbol{\sigma}^2 + (1 - \rho)\,\boldsymbol{s}^2,
\end{align}
where $\rho$ is the momentum parameter balancing historical and current statistics, $\bar{\boldsymbol{t}}$ is the element-wise mean vector of the current batch samples, and $\boldsymbol{s}^2$ is the corresponding element-wise variance vector. These updates are applied across all feature dimensions to keep the running statistics consistent with the evolving data distribution.

By maintaining domain-specific normalization parameters, this method allows each client model to preserve personalized feature characteristics while improving the stability and robustness of global clustering.

\subsubsection{Semantic Cluster}
After performing domain-aware normalization, the server further conducts global semantic modeling on the item text representations $\widetilde{\boldsymbol{t}}_i$ collected from different clients to learn shared semantic prototypes across domains. To achieve this, we adopt the weighted K-means method to cluster item representations. Compared with hierarchical and density-based clustering methods, K-means offers better computational efficiency and better scalability, and can produce stable and deterministic cluster assignments in large-scale high-dimensional scenarios. Therefore, it is well-suited for global semantic modeling in federated cross-domain recommendation. In our implementation, we follow the improved K-means algorithm proposed in~\citep{lu2025FFMSR}, which introduces a weighting mechanism to strengthen shared semantic features and thereby enhance semantic alignment across different domains.

The clustering process consists of three main steps:
\begin{enumerate}
    \item \textbf{Center Initialization:}
    We randomly select initial centers according to the predefined number of clusters $K$. 
    Let $\mathcal{D} = \{\mathcal{C}_1, \mathcal{C}_2, \ldots, \mathcal{C}_K\}$ denote the set of clusters and $\mathcal{S} = \{\boldsymbol{c}_1, \boldsymbol{c}_2, \ldots, \boldsymbol{c}_K\}$ denote the corresponding set of cluster centers. Each center $\boldsymbol{c}_k$ is initialized as:
\begin{equation}
    \boldsymbol{c}_k \sim \{\widetilde{\boldsymbol{t}}_1, \widetilde{\boldsymbol{t}}_2, \ldots, \widetilde{\boldsymbol{t}}_{|\mathcal{I}|}\}, 
    \quad k = 1, \ldots, K,
\end{equation}
where $\mathcal{I}$ denotes the global item set collected at the server from all clients.

 \item \textbf{Cluster Assignment:} For each data point $\widetilde{\boldsymbol{t}}_i$, we calculate its distance to all centers and assign it to the nearest one: 
    \begin{equation}
    \mathcal{C}_j = \left\{ \widetilde{\boldsymbol{t}}_i \;\middle|\; 
    j = \arg\min_{1 \le k \le K} \|\widetilde{\boldsymbol{t}}_i - \boldsymbol{c}_k\|^2 \right\}
    \end{equation}
   
    %
    
    \item \textbf{Center Update:} When updating the cluster centers, we adopt a weighted aggregation strategy. Specifically, the contribution of each data point to the center is determined by its distance, where a smaller distance yields a larger weight. The update formula is defined as:
    \begin{equation}
    \boldsymbol{c}_j = \frac{\sum_{\widetilde{\boldsymbol{t}}_i \in \mathcal{C}_j} w_i \widetilde{\boldsymbol{t}}_i}{\sum_{\widetilde{\boldsymbol{t}}_i \in \mathcal{C}_j} w_i},
    \end{equation}
where $w_i = \frac{1}{\|\widetilde{\boldsymbol{t}}_i - \boldsymbol{c}_j\| + \delta}$, and $\delta$ is a small constant to prevent numerical instability. Finally, the updated cluster centers $\mathcal{S}=\{\boldsymbol{c}_1,\boldsymbol{c}_2,\ldots,\boldsymbol{c}_K\}$ are sent back to all clients as global semantic knowledge.

\end{enumerate}

\subsection{Fine-Grained Semantic Adaptation and Transfer (FGSAT) Module}
The \ac{FGSAT} module is designed to enhance semantic knowledge transfer in federated cross-domain recommendation. Using the global semantic cluster centers generated on the server as semantic seeds, the module constructs fine-grained secondary semantic centers and builds a semantic graph together with local textual features. A \ac{GNN} is then employed to propagate information over the semantic graph, modeling the correlations between semantics and optimizing the representation learning process. Furthermore, the module incorporates an attention mechanism to achieve adaptive fusion of multi-source semantic information. Meanwhile, knowledge distillation and feature alignment constraints are introduced to improve the consistency of cross-domain semantic representations and enhance transfer performance.
Overall, by dynamically re-evaluating semantic centers and performing local adaptive adjustment, the \ac{FGSAT} module effectively alleviates cross-domain distribution discrepancies, thereby achieving more thorough semantic alignment and fusion.

\textbf{Semantic-Driven Two-Level Clustering.}
To improve the client’s adaptation to global semantics in cross-domain recommendation, we design a mechanism to construct fine-grained semantic centers based on the global cluster centers $\mathcal{S}$ delivered by the server. Specifically, let the local item text representations in the current batch be denoted as $\boldsymbol{T} \in \mathbb{R}^{B \times d_t}$, where $B$ denotes the batch size and each row $\boldsymbol{t}_i$ denotes the textual embedding of the item $i$. After receiving the global cluster centers $\mathcal{S}$, the client combines them with local text representations $\boldsymbol{T}$ and constructs the cluster indicator matrix $\boldsymbol{Y} \in \{0, 1\}^{B \times K}$ via one-hot encoding, where $\boldsymbol{Y}_{ik} = 1$ indicates that the item $i$ is assigned to the $k$-th cluster.

Then, the normalized secondary semantic centers are computed as:
\begin{equation}
   \boldsymbol{\tilde{C}} = (\boldsymbol{Y}^\top \boldsymbol{Y} + \eta \boldsymbol{I})^{-1} \boldsymbol{Y}^\top \boldsymbol{T},
\end{equation}
where $\eta$ is a small constant for numerical stability, and $\boldsymbol{I} \in \mathbb{R}^{K \times K}$ denotes the identity matrix, which is used to alleviate matrix singularity caused by sparse local sample distribution.

To further enhance the semantic expressiveness, a nonlinear transformation is applied to the secondary semantic centers:
\begin{equation}
    \boldsymbol{C}' = \operatorname{ReLU}(\boldsymbol{\tilde{C}} \boldsymbol{W}_1), 
\end{equation}
where $\boldsymbol{W}_1 \in \mathbb{R}^{d_t \times d_t}$ is a trainable parameter matrix. This projection maps the secondary semantic centers into a more discriminative representation space, thus strengthening their semantic modeling capability.

\textbf{Graph-Structured Knowledge Transfer.}  
Although the secondary semantic centers $\boldsymbol{C}'$ capture fine-grained local semantic features, fully leveraging their structural relationships with the original text representations $\boldsymbol{T}$ is still necessary to facilitate effective semantic information transfer. Following the graph-based relational knowledge transfer method proposed by~\citet{li2023SRTrans}, we stack them to form the node feature matrix $\boldsymbol{Z} = [\boldsymbol{C}'; \boldsymbol{T}] \in \mathbb{R}^{(K+B) \times d_t}$. The adjacency matrix $\boldsymbol{A}$ is constructed in a fully-connected manner by computing pairwise cosine similarity over $\boldsymbol{Z}$. Subsequently, a
\ac{GNN} is employed to propagate semantic information across the graph, with its core operation defined as:
\begin{align}
    \boldsymbol{L} &= \boldsymbol{D}^{-1/2} \boldsymbol{A} \boldsymbol{D}^{-1/2}, \\
    \boldsymbol{Z}' &= \operatorname{ReLU}(\boldsymbol{L} \boldsymbol{Z} \boldsymbol{W}_2),
\end{align}
where $\boldsymbol{D}$ is the degree matrix, $\boldsymbol{L}$ is the symmetric normalized adjacency matrix and $\boldsymbol{W}_2 \in \mathbb{R}^{d_t \times d_t}$ is a learnable parameter matrix for graph convolution. Through graph propagation, the model incorporates local adjacency information between nodes. Specifically, each enhanced item semantic representation $\boldsymbol{t}'_i$ (extracted from the nodes corresponding to the original items $\boldsymbol{T}$ in the updated matrix $\boldsymbol{Z}'$) improves its semantic context awareness by aggregating information from connected semantic centers and other items. This mechanism helps capture richer cross-domain semantic relationships, thus improving the semantic quality and generalization of the transferred representations.

\textbf{Attention-Based Fusion Mechanism.}
After obtaining the item semantic representation $\boldsymbol{t}_i'$ enhanced by graph-structured knowledge transfer, we adopt an attention-based dynamic fusion strategy to effectively integrate multi-level semantic information. Specifically, we consider three types of features: the transferred representation $\boldsymbol{t}_i'$ of item $i$, the center $\boldsymbol{c}_k$ of the $k$-th global cluster to which item $i$ belongs, and the center $\boldsymbol{c}_j'$ of the $j$-th secondary semantic cluster to which item $i$ belongs.

Each feature is first fed into a learnable scoring function \( \text{score}(\cdot) \), implemented as a trainable linear transformation layer:
$\text{score}(\boldsymbol{x}) = \boldsymbol{w}^\top \boldsymbol{x} + b$, where $\boldsymbol{w} \in \mathbb{R}^{d_t}$, $b \in \mathbb{R}$. This design maps the input semantic representation to a scalar score to evaluate the importance of each semantic component. 

The resulting scores are normalized via the Softmax function to obtain attention weights:
\begin{equation}
\boldsymbol{q}_i = \text{Softmax}\big( [\,
\text{score}(\boldsymbol{t}'_i),\ 
\text{score}(\boldsymbol{c}_k),\ 
\text{score}(\boldsymbol{c}'_j)
\,] \big),
\end{equation}
where \( \boldsymbol{q}_i \) denotes the dynamically computed attention weights for the three semantic components. 

The final fused representation is computed as:
\begin{equation}
\boldsymbol{t}_i'' = q_{i,1} \boldsymbol{t}_i' + q_{i,2} \boldsymbol{c}_k + q_{i,3} \boldsymbol{c}_j'.
\end{equation}

This mechanism enables the model to adaptively adjust the importance of different semantic sources for each sample, achieving effective multi-source semantic fusion. It enhances the expressiveness and adaptability of the transferred representation while alleviating the bias or information insufficiency caused by relying on a single semantic source.

\textbf{Knowledge Distillation and Feature Alignment.}  
This stage focuses on learning federated semantic representations for subsequent semantic transfer and reuse. To this end, we design two complementary objectives: knowledge distillation and feature alignment, which jointly ensure effective semantic transfer and stable model optimization.  

The knowledge distillation loss minimizes the Kullback-Leibler (KL) divergence between the original prediction $f(\boldsymbol{x}_u, \boldsymbol{t}_i)$ and the prediction based on the fused representation $f(\boldsymbol{x}_u, \boldsymbol{t}''_i)$:
\begin{equation}
    L_{\text{KD}} = \sum_{i=1}^{B} 
    \operatorname{KL}\!\big(f(\boldsymbol{x}_u, \boldsymbol{t}''_i)\, \|\, f(\boldsymbol{x}_u, \boldsymbol{t}_i)\big),
\end{equation}
where $f(\boldsymbol{x}_u, \boldsymbol{t}_i)$ and $f(\boldsymbol{x}_u, \boldsymbol{t}''_i)$ denote the predicted probability distributions using the original and fused representations, respectively.  
This objective enables global semantic knowledge transfer to local models by aligning their predictions. 
However, due to data heterogeneity across clients, relying solely on knowledge distillation may lead to inconsistent convergence behaviors, potentially causing representation shift and training instability.  

To mitigate these issues, we introduce a feature alignment loss that constrains the embedding-level consistency between the transferred and fused representations:
\begin{equation}
    L_{\text{FA}} = 
    \frac{1}{B} \sum_{i=1}^{B} 
    \|\boldsymbol{t}'_i - \boldsymbol{t}''_i\|_2^2.
\end{equation}
By enforcing proximity between the two representations in the feature space, this loss acts as a stabilizer during training, reducing client-level drift and improving cross-domain semantic coherence.  

The overall optimization objective is:
\begin{equation}
    L_{\text{total}} = 
    L_{\text{rec}} +
    \lambda_{\text{KD}} L_{\text{KD}} +
    \lambda_{\text{FA}} L_{\text{FA}},
\end{equation}
where $L_{\text{rec}}$ is the recommendation loss, and $\lambda_{\text{KD}}$ and $\lambda_{\text{FA}}$ are balancing coefficients.  
The total loss integrates three components:  
the recommendation loss $L_{\text{rec}}$ ensures effective modeling of local user-item interactions;  
the knowledge distillation loss $L_{\text{KD}}$ transfers global semantic knowledge to local models by minimizing prediction divergence;  
and the feature alignment loss $L_{\text{FA}}$ enhances stability and consistency by aligning fused representations with their transferred counterparts.  
Together, these losses enable accurate local recommendations while producing robust and transferable semantic representations for the next stage.

\subsection{Fine-Tuning on Local Clients}
After Stage 1, each domain generates pre-trained textual representations $\boldsymbol{t}_i^{\text{pre}}$ using the text encoder and organizes them into an embedding matrix $\boldsymbol{T}^{\text{pre}}$. These pre-trained representations serve as semantic priors and provide fundamental support for the local optimization in Stage 2. Unlike Stage 1, which mainly focuses on global semantic modeling, Stage 2 aims to further achieve semantic fusion and local guidance based on the pre-trained knowledge. Specifically, the model fuses the global semantic priors obtained from pre-training with the local data characteristics and personalized semantic information of each client, thereby improving personalized modeling capability while maintaining global consistency. This two-stage design of pre-training–then–fusion can effectively alleviate the negative transfer problem in federated learning and promote deeper semantic interaction and local adaptation, thus improving the model's robustness, training efficiency, and interpretability.

\textbf{Graph-Based Semantic Representation Learning.} 
We construct two types of semantic graphs with complementary properties: a local semantic graph and a pre-trained semantic graph. 

Following the semantic graph construction principle proposed by~\citet{zhang2022MICRO}, we construct the local semantic graph within each client based on cosine similarities between items. The local textual representations are learned by the local \ac{FGSAT} module, which fully exploits the semantic characteristics of local data and captures personalized item relationships. For each item $i$, its similarity to other items is calculated via the cosine function:
\begin{equation}
    s_{ij}^{\text{local}} = 
    \frac{(\boldsymbol{t}_i^{\text{local}})^\top \boldsymbol{t}_j^{\text{local}}}
    {\|\boldsymbol{t}_i^{\text{local}}\|_2 \, \|\boldsymbol{t}_j^{\text{local}}\|_2},
\end{equation}
where $\boldsymbol{t}_i^{\text{local}}$ and $\boldsymbol{t}_j^{\text{local}}$ denote the local textual representations of items $i$ and $j$, respectively, and $s_{ij}^{\text{local}}$ represents their semantic similarity.

To obtain a sparse and representative structure, we retain only the top-$N$ highest similarity connections for each item, forming the adjacency matrix $\boldsymbol{S}^{\text{local}}$. 

To stabilize graph learning, the adjacency matrix is normalized as follows:
\begin{equation}
    \tilde{\boldsymbol{S}}^{\text{local}} = 
    (\boldsymbol{D}^{\text{local}})^{-\tfrac{1}{2}} 
    \boldsymbol{S}^{\text{local}} 
    (\boldsymbol{D}^{\text{local}})^{-\tfrac{1}{2}},
\end{equation}
where $\boldsymbol{D}^{\text{local}}$ denotes the degree matrix with $D_{ii}^{\text{local}} = \sum_j s_{ij}^{\text{local}}$.

The pre-trained semantic graph is constructed in the same way as the local semantic graph, but is based on the semantic embeddings $\boldsymbol{T}^{\text{pre}}$ obtained from the pre-training stage. It incorporates shared semantic information across all clients to guide the local model in adapting to cross-domain semantic structures. The resulting $\tilde{\boldsymbol{S}}^{\text{local}}$ and $\tilde{\boldsymbol{S}}^{\text{pre}}$ are the normalized adjacency matrices of the local and pre-trained graphs, respectively, which are used in subsequent graph convolution propagation.

We then apply a \ac{GCN} to both the local and pre-learned semantic graphs for information propagation and feature fusion. The input embedding matrix $H_0$ is initialized with the same item ID embedding matrix $E$. Through this operation, item ID embeddings are propagated over the semantic graphs to absorb textual semantics, yielding deeply fused text-ID representations that tightly integrate structural interaction signals with semantic information. The graph convolution operations are defined as:
\begin{align}
    H_{\text{local}} &= \tilde{\boldsymbol{S}}^{\text{local}} H_0, \\ 
    H_{\text{pre}} &= \tilde{\boldsymbol{S}}^{\text{pre}} H_0,
\end{align}
where the $i$-th row $\boldsymbol{h}_i^{\text{local}}$ denotes the local fused representation of item $i$. 
Similarly, the $i$-th row of $H_{\text{pre}}$ represents the pre-learned fused representation $\boldsymbol{h}_i^{\text{pre}}$.

\textbf{Pre-trained–Local Contrastive Learning.}
To achieve deep semantic fusion between pre-trained and local representations, we introduce a bidirectional contrastive learning objective. 
This mechanism encourages consistency between semantically aligned item representations drawn from both the pre-trained and local spaces, while pushing apart unrelated ones. Such bidirectional supervision enables effective cross-space knowledge alignment and enhances representation robustness.

Given the pre-trained and local fused representations \( \boldsymbol{h}_i^{\text{pre}} \) and \( \boldsymbol{h}_i^{\text{local}} \), 
we first define the similarity function as:
\begin{equation}
\operatorname{sim}(\boldsymbol{h}_i, \boldsymbol{h}_j) = 
\frac{\boldsymbol{h}_i^\top \boldsymbol{h}_j}{\|\boldsymbol{h}_i\| \cdot \|\boldsymbol{h}_j\|}.
\end{equation}

Then, the bidirectional contrastive loss is formulated as:
\begin{align}
L_{\text{con}} = \sum_{i=1}^B \Bigg[
& -\log \frac{
\exp \big(\operatorname{sim}(\boldsymbol{h}_i^{\text{local}}, \boldsymbol{h}_i^{\text{pre}}) / \tau \big)
}{
\sum_{\substack{j=1 \\ j \neq i}}^B 
\exp \big(\operatorname{sim}(\boldsymbol{h}_i^{\text{local}}, \boldsymbol{h}_j^{\text{pre}}) / \tau \big)
} \nonumber \\
& -\log \frac{
\exp \big(\operatorname{sim}(\boldsymbol{h}_i^{\text{pre}}, \boldsymbol{h}_i^{\text{local}}) / \tau \big)
}{
\sum_{\substack{j=1 \\ j \neq i}}^B 
\exp \big(\operatorname{sim}(\boldsymbol{h}_i^{\text{pre}}, \boldsymbol{h}_j^{\text{local}}) / \tau \big)
}
\Bigg],
\end{align}
where \(\tau\) is the temperature parameter and $B$ denotes the batch size. This contrastive objective explicitly aligns the pre-trained and local semantic spaces, 
facilitating deep knowledge fusion and ensuring that the pre-trained semantic priors are effectively integrated into local representation learning.

\textbf{Multi-View Semantic Fusion.}
After obtaining the local fused representation and the pre-trained fused representation, 
we integrate multi-view semantic information to form a unified recommendation representation for each item:
\begin{equation}
\boldsymbol{x}_i = 
\boldsymbol{x}_i^{id} 
+ \frac{\boldsymbol{t}_i^{\text{local}}}{\|\boldsymbol{t}_i^{\text{local}}\|_2} 
+ \frac{\boldsymbol{h}_i^{\text{local}}}{\|\boldsymbol{h}_i^{\text{local}}\|_2} 
+ \frac{\boldsymbol{h}_i^{\text{pre}}}{\|\boldsymbol{h}_i^{\text{pre}}\|_2},
\end{equation}
where $\boldsymbol{x}_i^{id}$ denotes the item ID embedding, 
$\boldsymbol{t}_i^{\text{local}}$ represents the textual semantics learned from local data, 
$\boldsymbol{h}_i^{\text{local}}$ captures the local structure-aware features obtained via GCN propagation, 
and $\boldsymbol{h}_i^{\text{pre}}$ encodes the pre-trained semantic knowledge transferred from the federated stage. 

The final user preference score for item $i$ by user $u$ is computed as:
\begin{equation}
\hat{y}_{ui} = \boldsymbol{x}_i^\top \boldsymbol{x}_u,
\end{equation}
where $\boldsymbol{x}_u$ denotes the user embedding. 

Through this fusion mechanism, 
the model jointly leverages local semantics, structural dependencies, and pre-trained cross-domain knowledge, 
thereby generating recommendation representations that are both personalized and semantically generalizable.

\textbf{Optimization Objective.}
The overall optimization objective consists of three parts: 
the Bayesian Personalized Ranking (BPR) loss \(L_{\text{rec}}\), 
the pre-trained and local contrastive loss \(L_{\text{con}}\), 
and the local knowledge distillation loss \(L_{\text{kd}}\) computed based on the \ac{FGSAT} model:
\begin{equation}
L = L_{\text{rec}} + \alpha L_{\text{con}} + \beta L_{\text{kd}},
\end{equation}
where \(\alpha\) and \(\beta\) are hyperparameters that control the contributions of the contrastive and knowledge distillation losses, respectively. 
Specifically, \(\alpha\) governs the depth of semantic fusion between pre-trained and local representations, 
promoting meaningful interaction and integration of multi-level semantics, 
while \(\beta\) regulates the strength of local semantic adaptation guided by cluster-based knowledge. 
This joint optimization strategy enables the model to achieve deep fusion of pre-trained knowledge and local semantics, 
thereby improving both ranking accuracy and cross-domain generalization capability for personalized recommendation. The
details of the total framework are shown in Algorithm \ref{alg:fedcrf}.

\begin{algorithm}[t]
\caption{The Two-stage Training Process of FedCRF}
\label{alg:fedcrf}
\textbf{Input:} 
Item texts and user-item interactions from two domains $\mathrm{A}$ and $\mathrm{B}$; 
text encoder $\phi_{\text{emb}}$; 
cluster number $K$; 
training rounds $R$. \\
\textbf{Output:} 
Preference prediction $\hat{y}_{ui}$ for each user-item pair.
\begin{algorithmic}[1]

\STATE \textbf{Stage 1: Federated Pre-training}

\STATE \textbf{Initialization:} Initialize text encoder $\phi_{\text{emb}}$ on each client

\FOR{each communication round $r=1,2,\ldots,R$}
    \STATE \textbf{Client Executes:}
    \FOR{each client $j$}
        \FOR{each local batch}
            \STATE Encode item texts to obtain $\boldsymbol{t}_i=\phi_{\text{emb}}(\boldsymbol{x}_i^t)$
        \ENDFOR
        \STATE Upload normalized representations $\widetilde{\boldsymbol{t}}_i$ to server
    \ENDFOR

    \STATE \textbf{Server Executes:}
    \STATE Aggregate representations from all clients
    \STATE Perform semantic clustering and obtain global centers $\mathcal{S}=\{\boldsymbol{c}_k\}_{k=1}^K$
    \STATE Broadcast $\mathcal{S}$ and cluster assignments to all clients

    \STATE \textbf{Client Executes:}
    \FOR{each client $j$}
        \STATE Construct secondary semantic centers $\boldsymbol{C}'$ using $\mathcal{S}$ and local batch $\boldsymbol{T}$
        \STATE Build semantic graph with node features $\boldsymbol{Z}=[\boldsymbol{C}';\boldsymbol{T}]$
        \STATE Apply GNN propagation to obtain enhanced representations $\boldsymbol{t}'_i$
        \STATE Fuse multi-source semantics via attention to obtain $\boldsymbol{t}''_i$
        \STATE Update local encoder by minimizing $L_{\text{rec}} + \lambda_{\text{KD}} L_{\text{KD}} + \lambda_{\text{FA}} L_{\text{FA}}$
    \ENDFOR
\ENDFOR

\STATE Each client encodes all items and saves pre-trained semantic embeddings $\boldsymbol{T}^{\text{pre}}$

\STATE \textbf{Stage 2: Fine-tuning on Local Clients with Semantic Fusion}

\FOR{each client $j$}
    \STATE Construct local semantic graph $\tilde{\boldsymbol{S}}^{\text{local}}$ using $\boldsymbol{T}^{\text{local}}$
    \STATE Construct pre-trained semantic graph $\tilde{\boldsymbol{S}}^{\text{pre}}$ using $\boldsymbol{T}^{\text{pre}}$
    \STATE Perform graph convolution to obtain $\boldsymbol{h}_i^{\text{local}}$ and $\boldsymbol{h}_i^{\text{pre}}$
    \STATE Apply contrastive learning to minimize $L_{\text{con}}$
    \STATE Fuse ID and semantic representations to obtain $\boldsymbol{x}_i$
    \STATE Predict preference score $\hat{y}_{ui}=\boldsymbol{x}_u^\top \boldsymbol{x}_i$
    \STATE Optimize overall objective $L = L_{\text{rec}} + \alpha L_{\text{con}} + \beta L_{\text{kd}}$
\ENDFOR

\end{algorithmic}
\end{algorithm}

\section{Experiment setup \label{sec:experiment}}
This section outlines the key research questions investigated in our experiments, and describes the datasets, evaluation protocol, baseline methods for comparison, and implementation details of FedCRF in detail.

\subsection{Research Questions} 
We thoroughly assess our model by examining the following issues:
\begin{itemize}
    \item[\textbf{RQ1}] How does the proposed method perform compared with various state-of-the-art recommendation methods? 
    \item[\textbf{RQ2}] What are the individual contributions of the key components in FedCRF? 
    \item[\textbf{RQ3}] How sensitive is FedCRF to key hyper-parameters, including the contrastive loss weight, the knowledge distillation weight, and the number of global semantic clusters?
\end{itemize}

\subsection{Datasets}
We conduct experiments on three pairs of cross-domain datasets to evaluate our model, i.e., “Kitchen–Food”, “Care–Beauty”, and “Food–OnlineRetail”. Kitchen, Food, Care, and Beauty are all from Amazon~\citep{mcauley_amazon_2014}, which is a widely used real-world dataset in recommender systems, containing user reviews and metadata of products on the Amazon platform. The OnlineRetail dataset~\citep{chen_online_retail} is from an online retail company in the UK, containing user purchase records and brief descriptions of products. 

To assess the cross-domain performance of FedCRF, we first construct two cross-domain pairs from the Amazon dataset, namely “Kitchen–Food” and “Care–Beauty”. To further verify the generalization ability of the proposed method in the more challenging cross-platform scenario, we pair the OnlineRetail dataset with the Food dataset from Amazon to build a cross-platform and somewhat heterogeneous cross-domain recommendation scenario, denoted as “Food–OnlineRetail”. For this scenario, we conduct corresponding experiments and analyses, with detailed results shown in Section 6.5. To maintain the non-overlapping characteristic between domains, we do not perform any user alignment. Specifically, we do not match users across domains via identity information. Instead, we reassign user IDs so that they are numbered consecutively from 0. For item text in the Amazon dataset, we follow the standard practice in previous methods and use the product title, description, and brand as item text. In the OnlineRetail dataset, we directly use its default description as item text. To alleviate data sparsity, we adopt a five-core strategy that filters out users and items with fewer than five interactions. The statistics of the final constructed datasets are summarized in Table~\ref{tab:1}.

\begin{table}[h]
\centering
\captionsetup{justification=justified,singlelinecheck=false}
\caption{Statistics of the preprocessed datasets. ``Avg.~$n$'' denotes the average length of the user interaction sequence.}
\label{tab:1}  
\renewcommand{\arraystretch}{1.5}
\vspace{5pt}
\begin{tabular}{l r r r r}
\hline
Datasets & \#Users & \#Items & \#Inters. & Avg.~$n$ \\
\hline
Kitchen   & 18,028 & 9,560  & 137,799 & 7.64 \\
Food      & 14,066 & 8,272  & 145,127 & 10.32 \\
OnlineRetail    & 16,517 & 3,466  & 514,649 & 31.16 \\
\hline
Care      & 38,609 & 18,534 & 346,355 & 8.97 \\
Beauty    & 22,363 & 12,101 & 198,502 & 8.88 \\
\hline
\end{tabular}
\vspace{5pt}
\renewcommand{\arraystretch}{1}
\end{table}

\subsection{Evaluation Protocols}

In the evaluation phase, we split each user’s interaction history into training, validation, and test sets, with 80\% for training and 10\% for each of validation and test.
We treat each user-item interaction as a positive sample and adopt full ranking evaluation, where all items not interacted by the user are used as candidate negative items for ranking.
To measure recommendation performance, we employ two widely used evaluation metrics: Recall@K and NDCG@K, where K \(\in \{10, 20\} \).

\subsection{Baseline Methods}
To comprehensively verify the effectiveness of the proposed method, we conduct comparative experiments with several representative recommendation methods on multiple public datasets.
The selected baselines cover various paradigms, including single-domain graph neural recommendation, cross-domain recommendation, and federated cross-domain recommendation.

1) single-domain graph neural recommendation

\begin{itemize}
    \item \textbf{NGCF} \citep{wang2019neural}: A graph neural network-based collaborative filtering method that models user–item interactions as a bipartite graph. It captures high-order connectivity via multi-layer message propagation with feature transformation and non-linear activation, thereby enhancing the representation ability of users and items.

    \item \textbf{LightGCN} \citep{he2020lightgcn}: A simplified variant of NGCF that removes non-linear activation and feature transformation, retaining only neighbor aggregation to model high-order connectivity.

    \item \textbf{XSimGCL} \citep{yu2023xsimgcl}: A contrastive learning-based graph recommendation method that generates contrastive views through simple and efficient noise perturbation, avoiding complex graph augmentation and thus improving representation quality.
\end{itemize}

2) cross-domain recommendation
\begin{itemize}
    \item \textbf{GWCDR} \citep{li2022gromov}: It performs distribution-level alignment on representations from the source and target domains through a cross-domain alignment module to achieve knowledge transfer. It does not require sharing specific overlapping users or their behavior data across domains, thus better protecting user privacy.

    \item \textbf{SRTrans} \citep{li2023SRTrans}: Explores cross-domain associations through item clustering and semantic similarity, and propagates knowledge via the two-tier graph transfer network. It also alleviates negative transfer using task-driven knowledge distillation.

    \item \textbf{MITrans} \citep{li2024MITrans}: A cross-domain recommendation method based on mutual information constraints. It decouples domain-shared preferences and domain-specific preferences, and fuses shared preferences via a cross-domain graph structure to achieve knowledge transfer under non-overlapping user scenarios.
\end{itemize}

3) federated cross-domain recommendation
\begin{itemize}
    \item \textbf{FFMSR} \citep{lu2025FFMSR}: A federated semantic learning method for cross-domain recommendation constructs a multi-layer semantic encoder from raw item texts, performs clustering on the server to transfer semantic knowledge, and introduces an FFT-based filter and a gating mechanism to remove irrelevant semantic features. Since the method is designed for sequential recommendation, we adapt our datasets to a sequential format to ensure fair comparison.
    
    \item \textbf{FedDCSR} \citep{zhang2024feddcsr}: This approach aims to protect the data privacy of the domain, by introducing an inter-intra domain sequence representation disentanglement, which decomposes user sequence features into domain-shared and domain-exclusive types and learns global features across different domains through federated learning. Since the method is designed for sequential recommendation, we adapt our datasets to a sequential format to ensure fair comparison.

\end{itemize}

\subsection{Implementation Details}
FedCRF is implemented based on the PyTorch framework, and all experiments are conducted on an NVIDIA RTX 3090 GPU. The Adam optimizer is adopted for model training with a learning rate of 0.005 and a batch size of 1024.

In Stage 1, considering the scale differences among datasets, the number of global semantic clusters $K$ is set differently: 50 for Kitchen–Food and 70 for Care–Beauty. To achieve global semantic knowledge transfer and representation consistency optimization, we introduce the knowledge distillation loss $L_{\text{KD}}$ and feature alignment loss $L_{\text{FA}}$. Accordingly, their weight coefficients $(\lambda_{\text{KD}}, \lambda_{\text{FA})}$ are set to (0.2, 0.1) for Kitchen–Food and (0.1, 0.1) for Care–Beauty. Federated training is performed for 20 rounds.

In Stage 2, the temperature parameter $\tau$ is set to 0.5. The weight coefficients of the contrastive loss and knowledge distillation loss $(\alpha, \beta)$ are set to (0.2, 0.2) for Kitchen–Food and (0.1, 0.1) for Care–Beauty. To stabilize model convergence, a learning rate decay strategy and an early stopping mechanism are adopted. Meanwhile, the optimal model is selected based on the Recall@20 metric on the validation set for final testing.

All hyperparameters are tuned on the validation set to select the optimal configuration. Specifically, the number of global semantic clusters $K$ is chosen from the set $\{30, 40, 50, 60, 70\}$ and determined according to the Recall@20 metric. The loss weights $\alpha$, $\lambda_{\text{KD}}$, and $\lambda_{\text{FA}}$ are searched within the range $\{0.01, 0.1, 0.2, 0.3, 0.4\}$, and the best values are selected accordingly.

For all baseline models, we follow the hyperparameter settings in their original papers and conduct appropriate tuning on different datasets to ensure optimal performance.
To ensure experimental fairness, all baselines uniformly use the Adam optimizer with a learning rate of 0.005 and a batch size of 1024.
Specifically, for graph-based recommendation models, we uniformly adopt LightGCN as the basic architecture, with an embedding dimension of 64 and a network layer count of 2.
For models such as SRTrans, MITrans, and FFMSR, we strictly follow the settings in their original papers: we use the specified language models to generate text embeddings and retain their original clustering strategies. The number of cluster centers is kept consistent with that of our proposed method to ensure a fair comparison.
\section{Experimental Results (RQ1)}

\begin{table*}[ht]
\centering
\renewcommand{\arraystretch}{1.5}
\caption{Performance comparison of different methods on four datasets.}
\label{tab:results}
\resizebox{\textwidth}{!}{  
\begin{tabular}{llcccccccccc}
\hline
\textbf{ } & \textbf{ } & \multicolumn{3}{c}{\textbf{single-domain}} & \multicolumn{3}{c}{\textbf{cross-domain}} & \multicolumn{3}{c}{\textbf{federated}} & \textbf{Improv.} \\
\cmidrule(lr){3-5} \cmidrule(lr){6-8} \cmidrule(lr){9-11} \cmidrule(lr){12-12}
\textbf{Dataset} & \textbf{Metric} & \textbf{NGCF} & \textbf{LightGCN} & \textbf{XSimGCL} & \textbf{GWCDR} & \textbf{SRTrans} & \textbf{MITrans} & \textbf{FFMSR} & \textbf{FedDCSR} & \textbf{Ours} & \textbf{(\%)} \\
\hline
\multirow{4}{*}{Kitchen} 
& Recall@10 & 0.01749 & 0.03116 & 0.03032 & \underline{0.03572} & 0.03166 & 0.03159 & 0.02636 & 0.02572 & \textbf{0.04465*} & \textbf{22.2} \\
& NDCG@10 & 0.00951 & 0.01833 & 0.01674 & \underline{0.02137} & 0.01681 & 0.01700 & 0.01463 & 0.01421 & \textbf{0.02523*} & \textbf{12.8} \\
& Recall@20 & 0.02783 & 0.04549 & 0.04556 & \underline{0.05126} & 0.04983 & 0.04747 & 0.04373 &0.04281 & \textbf{0.06601*} & \textbf{22.4} \\
& NDCG@20 & 0.01227 & 0.02212 & 0.02081 & \underline{0.02563} & 0.02158 & 0.02122 & 0.01993 &0.01936 & \textbf{0.03099*} & \textbf{17.6} \\
\cmidrule(lr){2-12}
\multirow{4}{*}{Food} 
& Recall@10 & 0.05706 & 0.08191 & 0.08614 & \underline{0.08763} & 0.08354 & 0.08555 & 0.06939 & 0.06741 &\textbf{0.11078*} & \textbf{26.4} \\
& NDCG@10 & 0.03519 & 0.04992 & 0.05134 & \underline{0.05416} & 0.04853 & 0.04999 & 0.04358 &  0.04263 &\textbf{0.06708*} & \textbf{23.8} \\
& Recall@20 & 0.09263 & 0.12058 & 0.13049 & \underline{0.13358} & 0.13053 & 0.13008 & 0.12793 &  0.12486 & \textbf{0.16105*} & \textbf{20.5} \\
& NDCG@20 & 0.04584 & 0.06159 & 0.06472 & \underline{0.06855} & 0.06232 & 0.06307 & 0.06264 &  0.06112 & \textbf{0.08193*} & \textbf{19.5} \\
\hline
\multirow{4}{*}{Care} 
& Recall@10 & 0.03619 & 0.05981 & 0.05873 & \underline{0.06086} & 0.05511 & 0.05160 & 0.04763& 0.04602 & \textbf{0.07115*} & \textbf{16.9} \\
& NDCG@10 & 0.02060 & 0.03646 & 0.03627 & \underline{0.03634} & 0.03132 & 0.03001 & 0.02875 & 0.02802 & \textbf{0.04250*} & \textbf{16.9} \\
& Recall@20 & 0.05521 & 0.08370 & 0.08320 & \underline{0.08687} & 0.08072 & 0.07665 & 0.08411 & 0.08157 & \textbf{0.09897*} & \textbf{13.9} \\
& NDCG@20 & 0.02556 & 0.04327 & 0.04338 & \underline{0.04382} & 0.03865 & 0.03728 & 0.03953& 0.03891 & \textbf{0.05001*} & \textbf{14.1} \\
\cmidrule(lr){2-12}
\multirow{4}{*}{Beauty}
& Recall@10 & 0.05003 & 0.07704 & \underline{0.08034} & 0.07890 & 0.07281 & 0.07074 & 0.06557& 0.06321 & \textbf{0.09342*} & \textbf{16.3} \\
& NDCG@10 & 0.02818 & 0.04458 & \underline{0.04654} & 0.04640 & 0.03998 & 0.03994 & 0.03960 & 0.03842 & \textbf{0.05398*} & \textbf{15.9} \\
& Recall@20 & 0.07570 & 0.11159 & \underline{0.11884} & 0.11590 & 0.11101 & 0.10628 & 0.10502 & 0.10188 & \textbf{0.13384*} & \textbf{12.6} \\
& NDCG@20 & 0.03538 & 0.05420 & \underline{0.05738} & 0.05700 & 0.05420 & 0.04992 & 0.05146 & 0.05037 & \textbf{0.06513*} & \textbf{13.5} \\
\hline
\end{tabular}
}
\end{table*}

The results of the comparison across all datasets are shown in Table~\ref{tab:results}, leading to the following conclusions: 1) The proposed FedCRF outperforms all baseline methods across all datasets and evaluation metrics, demonstrating stable and consistent performance advantages. This shows that the method can effectively fuse ID and text modality information and achieve robust bridging of cross-domain semantic knowledge in non-overlapping scenarios. 2) Compared with single-domain methods (NGCF, LightGCN, XSimGCL), FedCRF achieves significant improvements. Single-domain methods only rely on local interaction data and struggle to alleviate data sparsity, whereas our method leverages cross-domain semantic information for knowledge transfer and representation enhancement. Although XSimGCL introduces contrastive learning to improve representation quality, its performance is still limited due to the lack of cross-domain information. 3) Compared with cross-domain methods (GWCDR, SRTrans, MITrans), FedCRF performs better. The ID-based method GWCDR relies on distribution-level alignment but is limited by discrete representations with insufficient semantic expressiveness. Text-based methods (SRTrans, MITrans) possess semantic modeling capabilities yet ignore ID-based collaborative information. In contrast, our method dynamically re-estimates semantic centers in the continuous semantic space and combines semantic graph modeling with ID-based collaborative information to achieve more effective cross-domain fusion. 4) Compared with federated methods (FFMSR, FedDCSR), FedCRF still exhibits clear advantages. Most existing methods are based on sequential modeling, which makes it difficult to capture high-order user–item structural relations. Moreover, their cross-domain fusion relies on static aggregation mechanisms, which tend to cause information loss and distribution shift.
\section{MODEL ANALYSIS\label{analysis}}
\subsection{Ablation Studies (RQ2)}

\begin{table*}[ht]
\renewcommand{\arraystretch}{1.5} 
\centering 
\caption{Ablation study of FedCRF on four domains}
\label{tab:ablation_1}
\vspace{1pt}

\resizebox{\linewidth}{!}{
\begin{tabular}{lcccccccc|cccccccc}
\toprule
\multirow{3}{*}{\textbf{Variants}} & \multicolumn{4}{c}{\textbf{Kitchen}} & \multicolumn{4}{c|}{\textbf{Food}} & \multicolumn{4}{c}{\textbf{Care}} & \multicolumn{4}{c}{\textbf{Beauty}} \\
\cmidrule(lr){2-17}
& \multicolumn{2}{c}{\textbf{Recall}} & \multicolumn{2}{c}{\textbf{NDCG}} & \multicolumn{2}{c}{\textbf{Recall}} & \multicolumn{2}{c|}{\textbf{NDCG}} & \multicolumn{2}{c}{\textbf{Recall}} & \multicolumn{2}{c}{\textbf{NDCG}} & \multicolumn{2}{c}{\textbf{Recall}} & \multicolumn{2}{c}{\textbf{NDCG}} \\
\cmidrule(lr){2-17} 
\cmidrule(lr){2-3} \cmidrule(lr){4-5} \cmidrule(lr){6-7} \cmidrule(lr){8-9} 
\cmidrule(lr){10-11} \cmidrule(lr){12-13} \cmidrule(lr){14-15} \cmidrule(lr){16-17}
& \textbf{@10} & \textbf{@20} & \textbf{@10} & \textbf{@20} & \textbf{@10} & \textbf{@20} & \textbf{@10} & \textbf{@20} & \textbf{@10} & \textbf{@20} & \textbf{@10} & \textbf{@20} & \textbf{@10} & \textbf{@20} & \textbf{@10} & \textbf{@20} \\
\midrule
FedCRF-Fed     & 0.04055 & 0.06144 & 0.02388 & 0.02934 & 0.10169 & 0.14975 & 0.06182 & 0.07607 & 0.06682 & 0.09423 & 0.03988 & 0.04757 & 0.09023 & 0.12894 & 0.05239 & 0.06311 \\
FedCRF-CL      & 0.04301 & 0.06217 & 0.02405 & 0.02878 & 0.10711 & 0.15607 & 0.06372 & 0.07786 & 0.06801 & 0.09564 & 0.04063 & 0.04829 & 0.09156 & 0.13109 & 0.05261 & 0.06348 \\
FedCRF-FGSAT   & 0.04359 & 0.06497 & 0.02492 & 0.03018 & 0.10894 & 0.15749 & 0.06481 & 0.07837 & 0.06891 & 0.09608 & 0.04117 & 0.04869 & 0.09004 & 0.13022 & 0.05259 & 0.06363 \\
\midrule
\textbf{FedCRF} 
& \textbf{0.04465} & \textbf{0.06601} & \textbf{0.02523} & \textbf{0.03112} & \textbf{0.11078} & \textbf{0.16105} & \textbf{0.06708} & \textbf{0.08193} & \textbf{0.07115} & \textbf{0.09897} & \textbf{0.04250} & \textbf{0.05001} & \textbf{0.09342} & \textbf{0.13384} & \textbf{0.05398} & \textbf{0.06513} \\
\bottomrule
\end{tabular}
} 
\vspace{1pt}
\end{table*}

To systematically verify the effectiveness of each core component in FedCRF, we conduct ablation experiments on the four domains: Kitchen, Food, Care, and Beauty, with results shown in Table~\ref{tab:ablation_1}.
The complete FedCRF framework consists of three core components: federated pre-training stage (Fed), client-side fine-grained semantic adaptation and transfer module (FGSAT), and bidirectional contrastive learning objective (CL) for aligning global and local representations.
As can be seen from the results, the full model achieves the best performance across all four domains, validating the effectiveness of the collaborative design of all modules. 1) In the FedCRF-FGSAT variant, the fine-grained semantic adaptation module is removed, and the client directly uses the server-side global cluster centers without local adaptive optimization.
Performance drops in all domains under this setting, indicating that static global clustering struggles to adapt to the local distribution of client data, easily introduces cross-domain semantic distribution shift, and thus weakens semantic alignment.
2) In the FedCRF-CL variant, the bidirectional contrastive learning objective is removed.
Although this setting still retains partial global semantic information, the lack of consistency constraints between global and local representations significantly degrades cross-domain semantic alignment, demonstrating the key role of contrastive learning in fine-grained semantic alignment.
3) Finally, in the FedCRF-Fed variant, the federated pre-training stage is completely removed, and each client trains the model independently.
This scheme yields the lowest performance among all settings, showing that without cross-domain semantic knowledge sharing, the model degenerates into single-domain learning, which can hardly alleviate data sparsity or effectively capture cross-domain commonalities. This further verifies the core role of federated semantic learning in knowledge transfer.

\subsection{Impact of Hyper-parameters (RQ3)}

\begin{figure*}
     \centering
     \includegraphics[width=15cm]{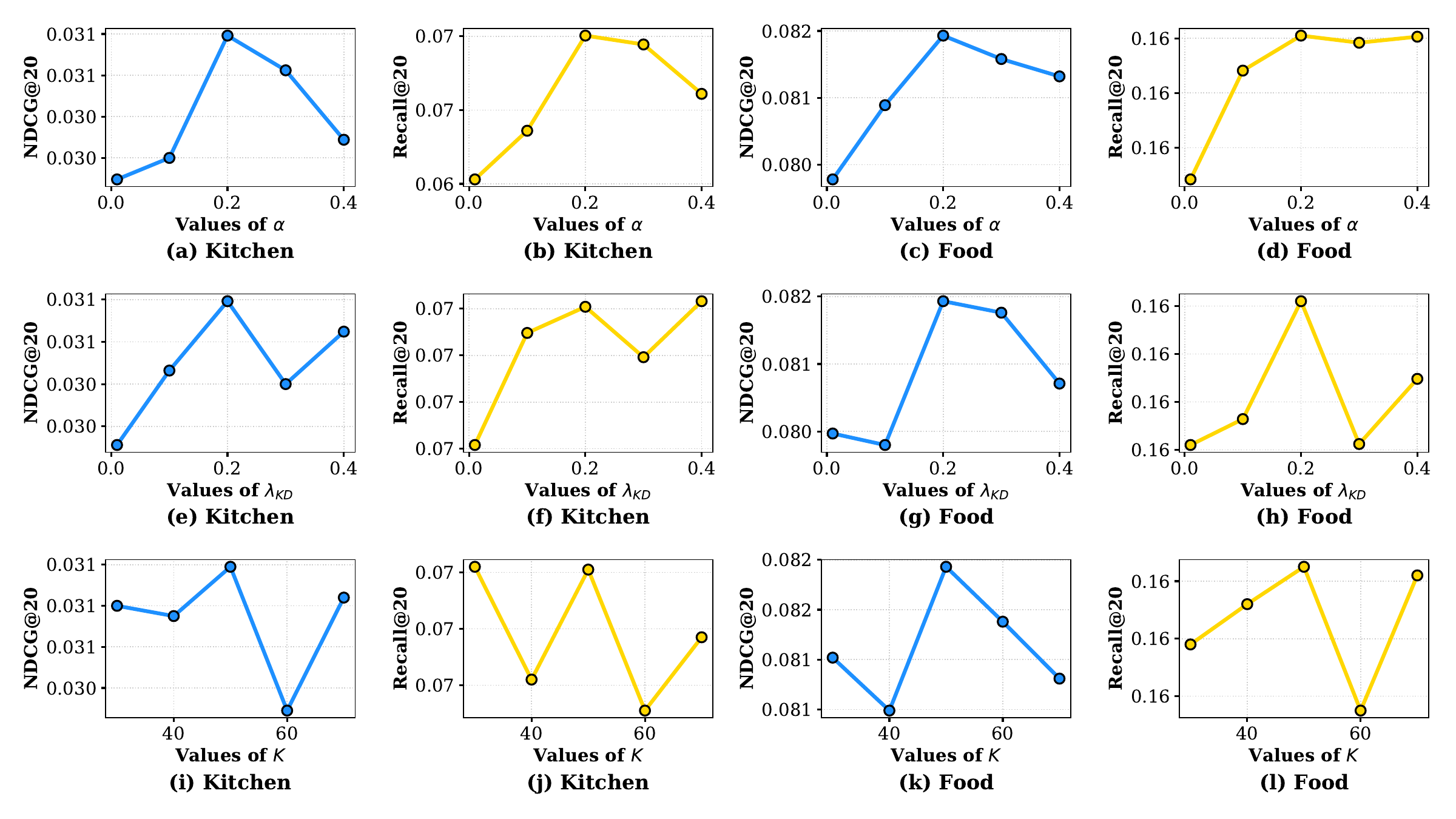}
     \caption{Performance sensitivity analysis with respect to three key hyperparameters: (a-d) contrastive loss weight $\alpha$, (e-h) knowledge distillation weight $\lambda_{\text{KD}}$, and (i-l) the number of global semantic clusters $K$.}
     \label{fig:hyper}
 \end{figure*}

To evaluate the sensitivity and stability of the model to key hyperparameters, we systematically analyze the effects of the contrastive loss weight $\alpha$, knowledge distillation weight $\lambda_{\mathrm{KD}}$, and the number of global semantic clusters $K$. 
As shown in Figure~\ref{fig:hyper}, we report the trends of NDCG@20 and Recall@20 on the Kitchen and Food domains, respectively.

\textbf{(1) Contrastive loss weight $\alpha$.}
As $\alpha$ gradually increases, the overall model performance first rises and then declines, reaching optimal or near-optimal performance around $\alpha = 0.2$. This phenomenon indicates that appropriate contrastive supervision can effectively strengthen the consistency between pre-trained global representations and local semantic representations. However, when $\alpha$ is excessively large, contrastive constraints dominate and suppress personalized representation learning, leading to performance degradation.

\textbf{(2) Knowledge distillation weight $\lambda_{\mathrm{KD}}$.}
The model performs relatively stably under different values of $\lambda_{\mathrm{KD}}$, demonstrating strong robustness to this hyperparameter. In most settings, the best or superior performance is achieved at $\lambda_{\mathrm{KD}} = 0.2$. This shows that, in the federated pre-training stage, proper knowledge distillation can effectively promote the transfer of global semantic knowledge while avoiding excessive interference with local personalized features.

\textbf{(3) The number of global semantic clusters $K$.}
Model performance exhibits a clear non-monotonic trend with varying $K$, achieving the best performance at a medium scale (e.g., $K = 50$). A small $K$ leads to overly coarse semantic partitioning, which limits representation capacity. Conversely, an excessively large $K$ may introduce semantic fragmentation and noise, weakening the generalization ability of cluster centers. These results suggest that properly controlling semantic granularity is crucial for effective cross-domain knowledge modeling.

\subsection{Comparison Results on Cross-Platform Scenario}

\begin{table*}[ht]
\renewcommand{\arraystretch}{1.2} 
\centering 
\caption{Performance comparison of different methods on Food and OnlineRetail domains.}
\label{tab:cross_platform}
\vspace{1pt}

\resizebox{\linewidth}{!}{
\begin{tabular}{lcccccccc}
\toprule
\multirow{3}{*}{\textbf{Variants}} & \multicolumn{4}{c}{\textbf{Food}} & \multicolumn{4}{c}{\textbf{OnlineRetail}} \\
\cmidrule(lr){2-9}
& \multicolumn{2}{c}{\textbf{Recall}} & \multicolumn{2}{c}{\textbf{NDCG}} & \multicolumn{2}{c}{\textbf{Recall}} & \multicolumn{2}{c}{\textbf{NDCG}} \\
\cmidrule(lr){2-9}
& \textbf{@10} & \textbf{@20} & \textbf{@10} & \textbf{@20} & \textbf{@10} & \textbf{@20} & \textbf{@10} & \textbf{@20} \\
\midrule
NGCF      & 0.01749 & 0.02783 & 0.00951 & 0.01227 & 0.16147 & 0.22961 & 0.13282 & 0.16482 \\
LightGCN      & 0.08191 & 0.12058 & 0.04992 & 0.06159 & 0.21805 & 0.30070 & 0.18208 & 0.21982 \\
XSimGCL       & 0.03032  & 0.04556  & 0.01674 & 0.02081 & 0.22377 & 0.30931 & 0.18785 & 0.22951 \\
GWCDR      
& 0.08872 & 0.12635 & 0.04791 & 0.06102 
& 0.21884 & 0.30615 & 0.17963 & 0.21767 \\
SRTrans       & 0.08111 &  0.12591 & 0.04868 &0.06173 & 0.19818 & 0.28043 & 0.16253 & 0.20075 \\
MITrans       & 0.06840 & 0.10284 & 0.04174 & 0.05219 & 0.19499 & 0.27551 & 0.16212 & 0.19896 \\
FFMSR      
& 0.08436 & 0.12318 & 0.04712 & 0.06344 
& 0.20142 & 0.28491 & 0.16573 & 0.20384 \\
FedDCSR      
& 0.08194 & 0.11976 & 0.04837 & 0.06102 
& 0.19685 & 0.27866 & 0.16104 & 0.19821 \\
\cmidrule(lr){1-9}
FedCRF-Fed     &0.10184& 0.14898& 0.06157& 0.07554 & 0.22108 & 0.30757 & 0.18312& 0.22291 \\
FedCRF-CL      & 0.10727 & 0.15666 & 0.06454 & 0.07894 & 0.22322 & 0.31066 & 0.18500 & 0.22416 \\
FedCRF-FGSAT   & 0.10682 & 0.15854 & 0.06359 & 0.07851 & 0.22401 & 0.30943 & 0.18330 & 0.22242 \\
\midrule
\textbf{FedCRF} & \textbf{0.11142} & \textbf{0.16185} & \textbf{0.06713} & \textbf{0.08189} & \textbf{0.22864} & \textbf{0.31488} & \textbf{0.19232} & \textbf{0.23368} \\
\bottomrule
\end{tabular}
}
\vspace{1pt}
\end{table*}

To further evaluate the capability of our method in cross-platform scenarios, we further conduct experiments on the OnlineRetail-Food dataset, and show the results in Table~\ref{tab:cross_platform}. From the table, we can have the following observations: 1) The proposed FedCRF method consistently outperforms all baselines on all metrics. This shows that even in cross-platform scenarios with heterogeneous data distributions, our two-stage FedCRF framework can effectively capture and transfer shared semantic knowledge across domains. Meanwhile, the model preserves domain-specific features through fine-grained local modeling, verifying its strong cross-platform generalization ability.
2) The cross-platform scenario significantly impacts the performance of the baselines.
Among single-domain methods, XSimGCL improves representation via contrastive learning and performs well on some platforms. However, due to the lack of cross-domain modeling, its performance fluctuates significantly across platforms. Among cross-domain methods, GWCDR has certain advantages in some scenarios, indicating that ID-based explicit alignment can alleviate cross-platform data isolation to some extent, but its overall performance is still unstable. In contrast, the federated method FFMSR performs more steadily across platforms, showing the benefits of cross-domain information aggregation. However, due to the lack of fine-grained semantic modeling, its overall performance is not optimal.
3) We also conduct ablation studies on these two platforms (i.e., FedCRF-Fed, FedCRF-CL, and FedCRF-FGSAT).
From the results in Table~\ref{tab:cross_platform}, we can find that our cross-domain semantic transfer and fusion modules are important to the FedCRF method. Without them, the performance of FedCRF is significantly reduced.

\subsection{Time Complexity and Runtime Analysis}

\begin{table}[t]
\centering
\caption{Time Complexity and Runtime Comparison on the Kitchen-Food Task (Kitchen Domain)}
\label{tab:time_complexity}
\small
\setlength{\tabcolsep}{18pt}
\setlength{\arrayrulewidth}{0.5pt}
\renewcommand{\arraystretch}{1.2}

\begin{tabular}{lcc}
\hline
\textbf{Algorithm} & \textbf{Time Complexity} & \textbf{Time per Epoch (Kitchen)} \\
\hline
GWCDR   
& $\mathcal{O}\!\left(N_s^{2}d + N_t^{2}d\right) 
 + \mathcal{O}\!\left(T_g N_s N_t\right)$ 
& $\sim$13s \\

SRTrans 
& $\mathcal{O}\!\left((M_s^{2}+M_t^{2})d\right) 
 + \mathcal{O}\!\left(H(E_s+E_t)d\right)$ 
& $\sim$12s \\

MITrans 
& $\mathcal{O}\!\left(K^{2}d\right) 
 + \mathcal{O}\!\left(K M d\right)
 + \mathcal{O}\!\left(H E d\right)$
& $\sim$15s \\

FFMSR 
& $\mathcal{O}\!\left(T_k M K d\right) 
 + \sum_{j=1}^{J}\mathcal{O}\!\left(M_j\,k\,d^{2}\right)$ 
&  $\sim$17s\\

FedDCSR
& $\sum_{j=1}^{J}\mathcal{O}\!\left(H E_j d + B T^{2} d + B^{2} d\right)$
&  $\sim$19s\\

\hline
FedCRF (Ours) 
& $\mathcal{O}\!\left(T_k M K d\right) 
 + \sum_{j=1}^{J}\mathcal{O}\!\left(M_j^{2}d + H E_j d\right)$ 
& $\sim$11s \\
\hline
\end{tabular}

\vspace{5pt}
\footnotesize
\parbox{\linewidth}{
\textbf{Note.}
The batch size $B$ is set to 1024. $N_s$ and $N_t$ denote the number of users in the source and target domains, respectively. $M_s$ and $M_t$ represent the number of items in the source and target domains, respectively. $M_j$ denotes the number of items on client $j$, with $M = \sum_{j=1}^{J} M_j$, where $J$ is the number of clients (domains). $d$ is the embedding dimension. $H$ denotes the number of graph propagation layers. $E_s$, $E_t$ (or $E_j$) represent the number of edges in the constructed graph. $K$ is the number of semantic clusters. $k$ denotes the dimension of the intermediate semantic projection space ($k \ll d$). $T_k$ is the number of clustering iterations. $T_g$ is the number of GW optimization iterations. $T$ is the sequence length.
}
\end{table}

Table~\ref{tab:time_complexity} compares the theoretical time complexity and practical training runtime of FedCRF with several competitive cross-domain recommendation methods.

In terms of theoretical complexity, the computational overhead of different methods mainly comes from three types of operations:
1) pairwise matching or alignment computation between cross-domain entities (e.g., $N_s N_t$ in GWCDR),
2) high-order semantic or sequence-based modeling (e.g., SRTrans, MITrans, and FFMSR), and
3) self-attention computation in sequential modeling (e.g., FedDCSR).
These operations typically introduce high computational overhead as the data scale or sequence length increases. In contrast, the computation of FedCRF mainly consists of server-side semantic clustering and client-side graph propagation, which avoids explicit full cross-domain matching and complex sequential modeling, thus achieving higher efficiency in its computational structure.

For actual running time, we record the average training time per epoch on the Kitchen domain in the Kitchen--Food task under the same experimental environment.
As shown in Table~\ref{tab:time_complexity}, although different methods differ in theoretical complexity, their running time exhibits a consistent trend with their computational structures: models involving complex cross-domain alignment or sequential modeling usually cost more time, while graph-based models achieve better efficiency.
FedCRF achieves the shortest training time among all methods, demonstrating its favorable computational efficiency and scalability while maintaining competitive performance.

\subsection{Visualization of Learned Representations}

\begin{figure}
  \centering
  \begin{minipage}{0.32\textwidth}
    \centering
    \includegraphics[width=1\linewidth]{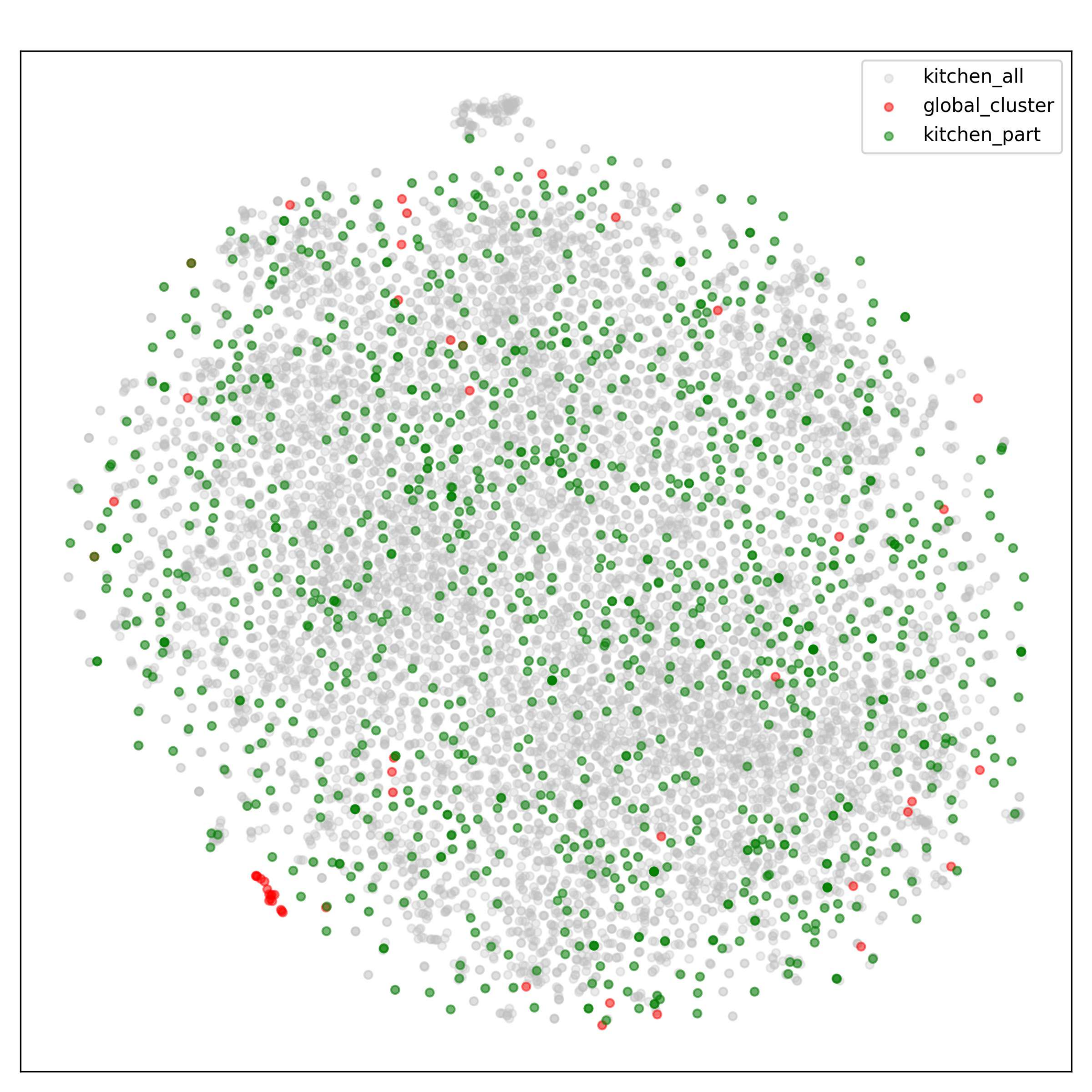}
    \centerline{\fontsize{5}{1}\selectfont(a) Global Clustering in Kitchen Domain} 
  \end{minipage}
  \hspace{2pt}
  \begin{minipage}{0.32\textwidth}
    \centering
    \includegraphics[width=1\linewidth]{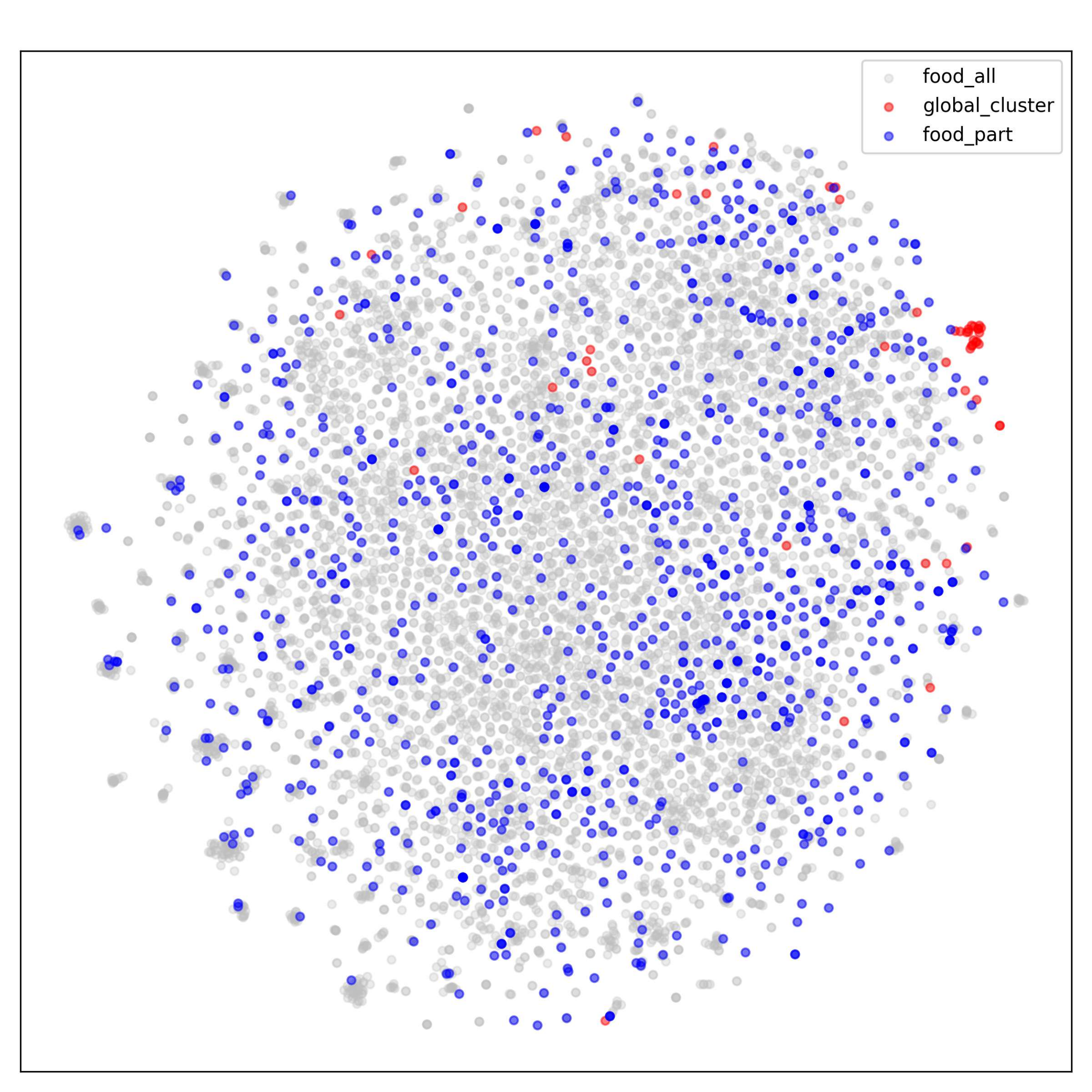}
    \centerline{\fontsize{5}{1}\selectfont(b) Global Clustering in Food Domain}
  \end{minipage}
  \hspace{2pt}
  \begin{minipage}{0.32\textwidth}
    \centering
    \includegraphics[width=1\linewidth]{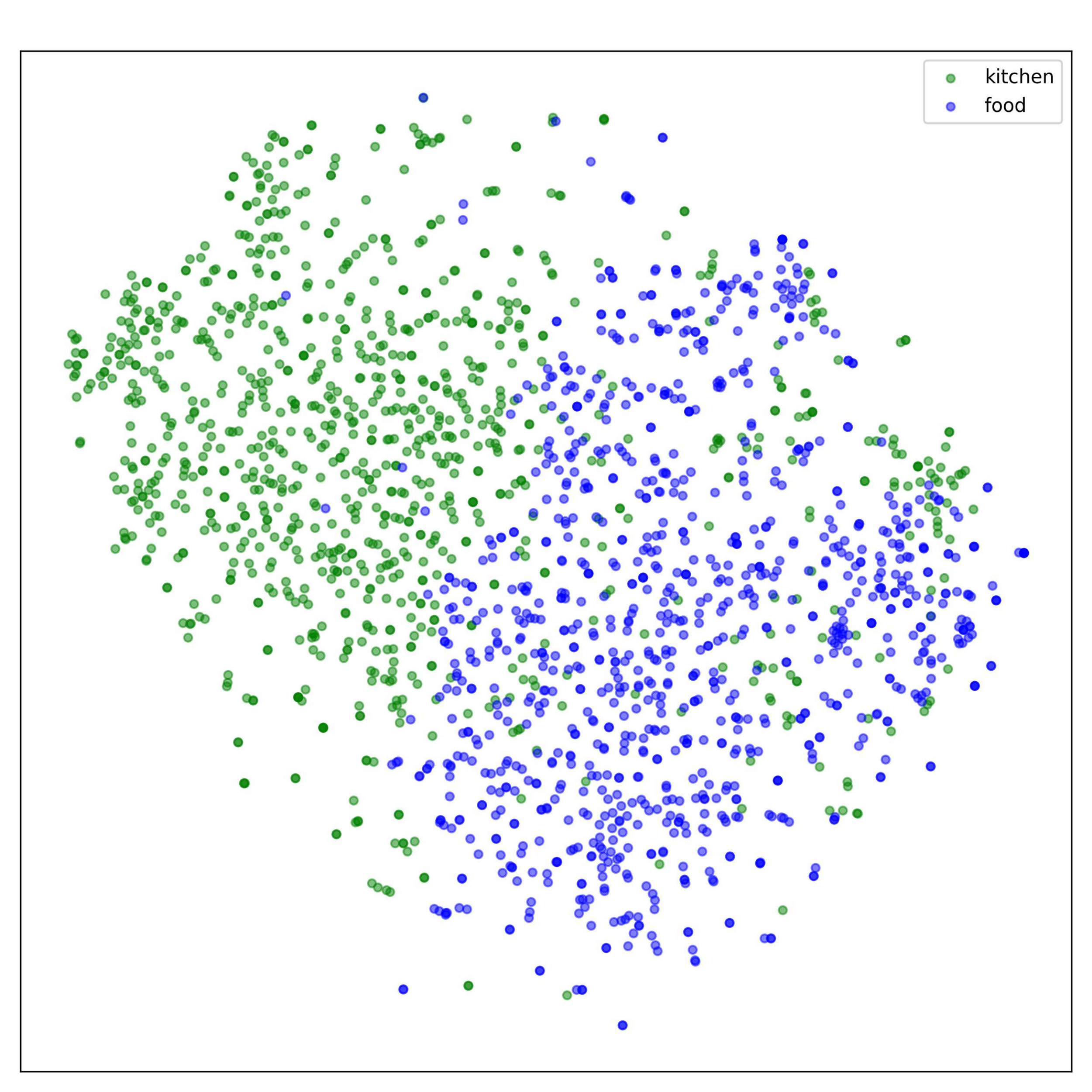}
    \centerline{\fontsize{5}{1}\selectfont(c) Embedding Distribution without FGSAT}
  \end{minipage}

  \vspace{0.2cm}

  \begin{minipage}{0.32\textwidth}
    \centering
    \includegraphics[width=1\linewidth]{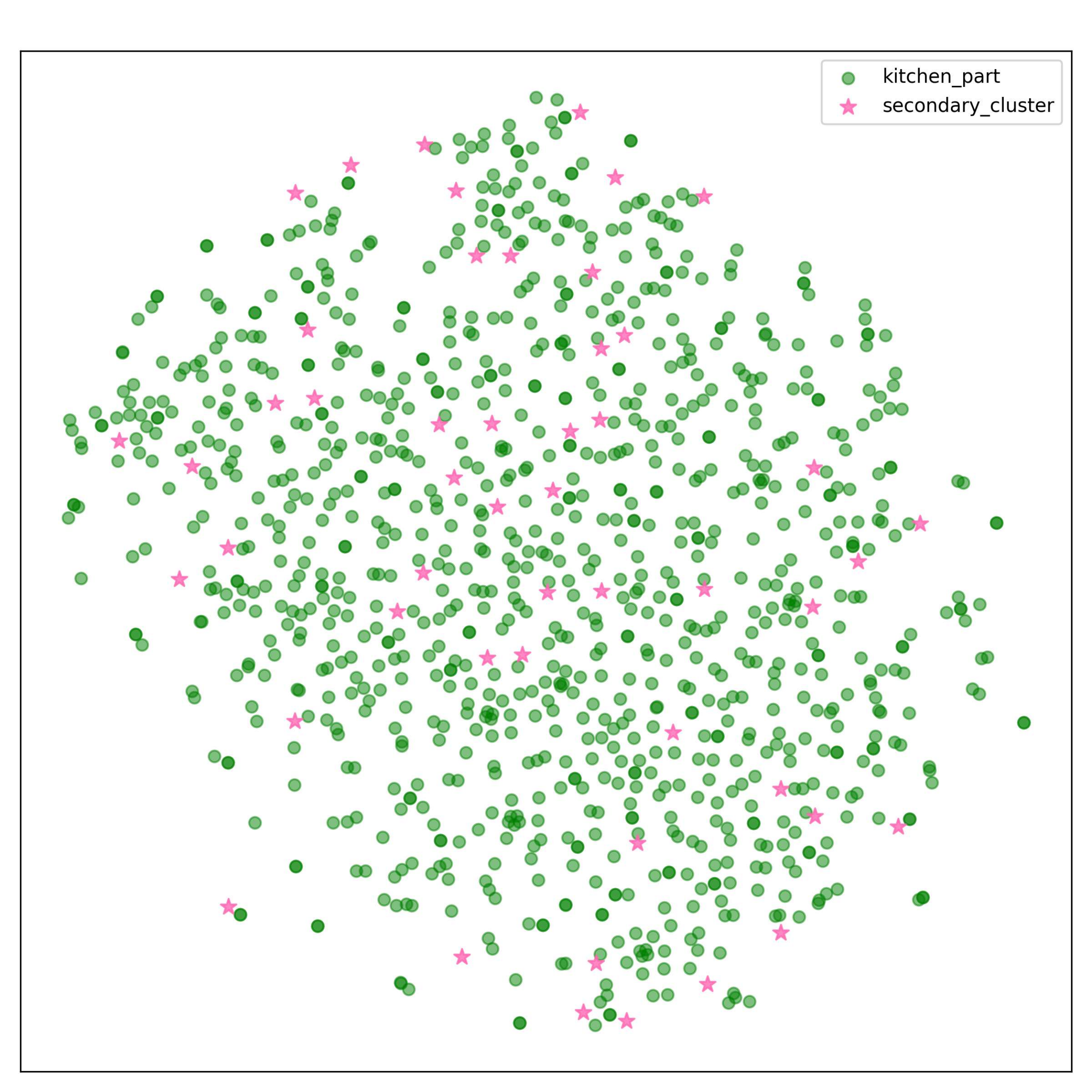}
    \centerline{\fontsize{5}{9}\selectfont(d) Secondary Clustering via FGSAT in Kitchen Domain}
  \end{minipage}
  \hspace{2pt}
  \begin{minipage}{0.32\textwidth}
    \centering
    \includegraphics[width=1\linewidth]{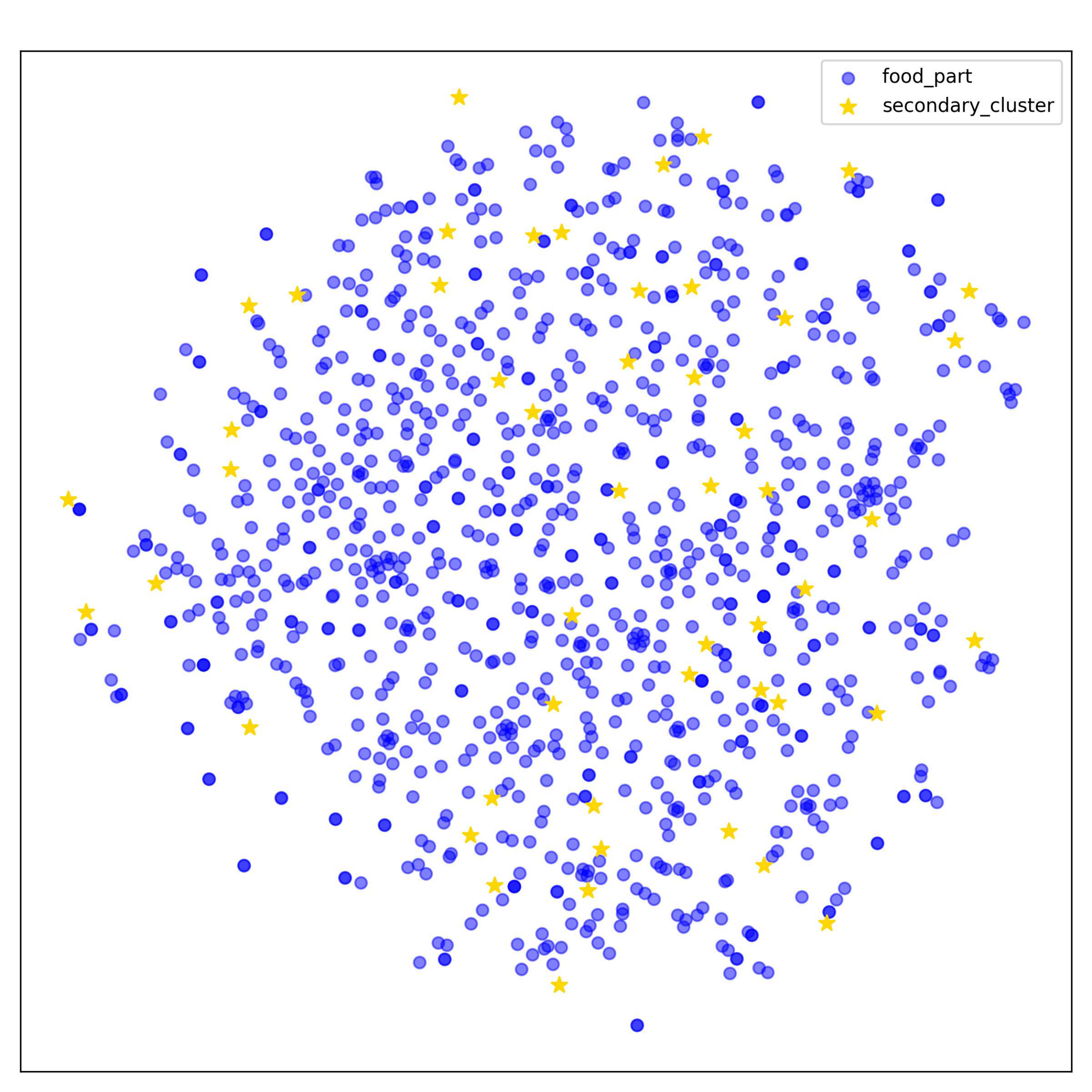}
    \centerline{\fontsize{5}{9}\selectfont(e) Secondary Clustering via FGSAT in Food Domain}
  \end{minipage}
  \hspace{2pt}
  \begin{minipage}{0.32\textwidth}
    \centering
    \includegraphics[width=1\linewidth]{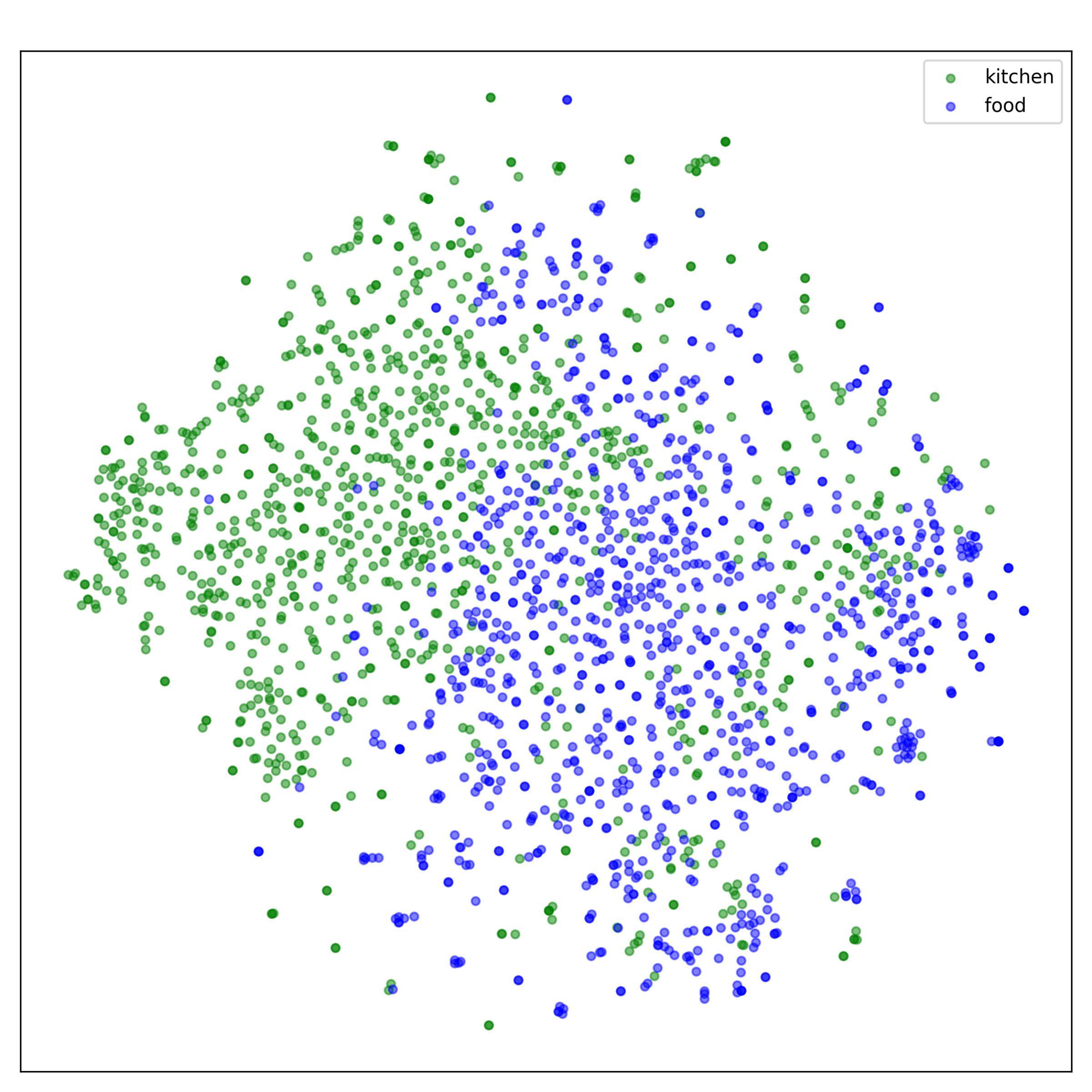}
    \centerline{\fontsize{5}{9}\selectfont(f) Embedding Distribution with FGSAT}
  \end{minipage}

  \caption{Visualization of Cross-domain Semantic Alignment across the Kitchen and Food domains}
  \label{fig:cross_domain_tsne}
\end{figure}

To verify the effectiveness of the proposed method in cross-domain semantic aggregation, we conduct visualization experiments via t-SNE~\citep{van2008tSNE}
on the item representations learned during the federated pre-training stage. 
Take the Kitchen and Food domain pair as an example,
we visualize the distribution of item representations with or without our  FGSAT module in Figure~\ref{fig:cross_domain_tsne}.
Figures~\ref{fig:cross_domain_tsne} (a), ~\ref{fig:cross_domain_tsne} (b), and~\ref{fig:cross_domain_tsne} (c) are the results without our FGSAT module. They only use the global clustering method for representation learning.
Figures~\ref{fig:cross_domain_tsne} (d),~\ref{fig:cross_domain_tsne} (e), and~\ref{fig:cross_domain_tsne} (f) are the results with our FGSAT module.

As shown in Figures~\ref{fig:cross_domain_tsne} (a) and~\ref{fig:cross_domain_tsne} (b), with global clustering, the cluster centers learned at the server can provide shared semantic guidance for data from different domains. These centers contain common cross-domain semantic information. However, these cluster centers mainly reflect common cross-domain features and do not fully fit the local data distribution of each client. As a result, many local samples are still scattered around or even far from the cluster centers in the semantic space. This shows that global clustering, as a static and global representation, only provides coarse-grained semantic alignment. It cannot well capture the local distribution of client data, which limits the fine-grained performance of cross-domain semantic aggregation.
We sample some instances from the two domains separately in Figures~\ref{fig:cross_domain_tsne} (a) and~\ref{fig:cross_domain_tsne} (b), and visualize them jointly in a unified semantic space to obtain Figure~\ref{fig:cross_domain_tsne} (c), so as to observe the cross-domain alignment effect more intuitively. It can be seen that without FGSAT, although the global semantic centers achieve cross-domain alignment to some extent, samples from different domains still show obvious distribution separation, with only limited overlap in local regions.This indicates that when relying only on static global clustering, the model can achieve preliminary semantic alignment, but cannot fully eliminate the distribution gap between different domains, resulting in still limited alignment performance.

The results that exploit our \ac{FGSAT} module are shown in Figures~\ref{fig:cross_domain_tsne} (d) and~\ref{fig:cross_domain_tsne} (e), from which we can observe that: 1) adaptively re-estimating semantic centers via our improved clustering method makes the semantic prototypes better fit the local data distribution, showing clear fine-grained semantic modeling ability. 
2) Besides, we also sample corresponding instances from the two domains in Figures~\ref{fig:cross_domain_tsne} (d) and~\ref{fig:cross_domain_tsne} (e) and visualize them jointly to obtain Figure~\ref{fig:cross_domain_tsne} (f), so as to analyze the cross-domain alignment effect after introducing \ac{FGSAT}. Compared with Figure ~\ref{fig:cross_domain_tsne} (c), we can clearly find that the sample distributions of different domains become more continuous, and
the degree of overlap is significantly improved, 
thus forming a more unified semantic space. 
This result demonstrates that our FGSAT method can effectively alleviate the cross-domain distribution shift via dynamically re-evaluating and locally adapting semantic centers, thereby achieving more sufficient cross-domain semantic alignment and fusion. 

\subsection{Analysis of Privacy Risks and Assumptions}

Although our federated framework only uploads item embeddings and keeps user interaction data on local clients, these embeddings may still contain implicit information about user behaviors, leading to potential privacy risks. To evaluate this risk, we adopt the Similarity-based Inference Attack (SIA)\citep{yuan2025fellas}. 
Specifically, we assume an attacker can intercept the uploaded item embeddings, and the attacker tends to infer other items that are interacted by the same user. 

For any intercepted item, we first retrieve the most similar Top-K items via computing the similarities between it and all the other items in the item set.
Then, we estimate the potential interacting users of the target item using the interaction data of these similar items (whether they are interacted by the same user).
Finally, we compare the predicted user set with the ground-truth set, and use F1 score to measure attack performance and quantify privacy leakage risk.

The experimental results are shown in Table ~\ref{tab:Privacy_Risk}, from which we can observe that: 1) On all datasets, the overall F1 scores are very low, meaning the attacker cannot easily recover user interactions using only item embeddings, and our model can provide effective privacy protection on users. 2) From Table ~\ref{tab:Privacy_Risk}, we can also observe that as Top-K increases, the attack performance consistently drops. This shows that adding more similar items may introduce extra noise, weakening the attacker’s ability to infer the target user set. 
3) The attack performance differs across different datasets. The attack achieves relatively higher performance in the Food–OnlineRetail dataset, while the performance is generally lower in Kitchen–Food and Care–Beauty. This result indicates that the user-item interaction distribution of each dataset has a significant impact on the SIA method. When similar items share more overlapping users, the attacker can more easily recover the target item’s user set.

\begin{table}[htbp]
\centering
\caption{The attack performance (F1 scores) for different Top-K settings on three datasets. Lower values indicate better data privacy protection abilities.}
\label{tab:Privacy_Risk}
\begin{tabular}{c|cc|cc|cc}
\hline
\textbf{Top-K} & \textbf{Kitchen} & \textbf{Food} & \textbf{Care} & \textbf{Beauty} & \textbf{Food} & \textbf{OnlineRetail} \\
\hline
1 & 0.0133 & 0.0335 & 0.0116 & 0.0182 & 0.0311 & 0.0538 \\
3 & 0.0123 & 0.0304 & 0.0101 & 0.0162 & 0.0271 & 0.0492 \\
5 & 0.0107 & 0.0253 & 0.0086 & 0.0138 & 0.0223 & 0.0436 \\
\hline
\end{tabular}
\end{table}

\section{{Conclusion\label{sec:conclusion}}}

This paper investigates \ac{PPCDR} under the fully non-overlapping setting. To address the limitations of existing methods in semantic modeling and global–local information fusion, we propose FedCRF. Specifically, we leverage textual semantics as a cross-domain bridge to enable knowledge transfer within a federated learning framework. On the server side, global semantic clustering is employed to capture shared semantic information, while on the client side, the \ac{FGSAT} module is designed to achieve dynamic adaptation and fine-grained modeling of global and local semantics. In addition, semantic graph modeling and contrastive learning are introduced to enable deep multi-source semantic fusion, and ID representations are further integrated to enhance recommendation performance. Experimental results on multiple real-world datasets demonstrate that FedCRF consistently outperforms existing methods.

This work has several limitations: 1) The proposed method mainly relies on textual semantic information, and its performance may be limited when textual data is sparse or of low quality. 2) The global clustering and semantic graph construction introduce additional computational overhead, which may affect scalability in large-scale scenarios. 3) In cross-platform scenarios, due to significant semantic distribution discrepancies across platforms, the performance improvement is relatively limited, leaving room for further improvement.

In the future, we will explore more efficient semantic modeling strategies, improve generalization under cross-platform heterogeneous settings, and extend the framework to multi-modal recommendation scenarios.

\section*{Data availability}
The data is publicly accessible.

\bibliographystyle{mymodel5}
\bibliography{cas-refs}

\end{document}